\documentclass[sigconf]{acmart}

\usepackage{booktabs} 
\usepackage{mathtools}
\usepackage{bm}
\usepackage{amsmath}
\usepackage{balance}
\usepackage{enumitem}

\newcommand{\xhdr}[1]{\vspace{2mm}\noindent{{\bf #1}}}

\setcopyright{iw3c2w3}

\begin{document}

\title{A Dynamic Embedding Model of the Media Landscape}

\author{J\'er\'emie Rappaz}
\authornote{Both authors contributed equally to the paper}
\affiliation{%
  \institution{EPFL}
  \institution{Switzerland}
}
\email{jeremie.rappaz@epfl.ch}

\author{Dylan Bourgeois}
\authornotemark[1]
\affiliation{%
  \institution{EPFL}
  \institution{Switzerland}
}
\email{contact@dtsbourg.me} 

\author{Karl Aberer}
\affiliation{%
  \institution{EPFL}
  \institution{Switzerland}
}
\email{karl.aberer@epfl.ch}

\copyrightyear{2019}
\acmYear{2019}
\acmConference[WWW '19]{Proceedings of the 2019 World Wide Web Conference}{May 13--17, 2019}{San Francisco, CA, USA}
\acmBooktitle{Proceedings of the 2019 World Wide Web Conference (WWW '19), May 13--17, 2019, San Francisco, CA, USA}
\acmPrice{}
\acmDOI{10.1145/3308558.3313526}
\acmISBN{978-1-4503-6674-8/19/05}

\begin{abstract}

Information about world events is disseminated through a wide variety of news channels, each with specific considerations in the choice of their reporting. Although the multiplicity of these outlets should ensure a variety of viewpoints, recent reports suggest that the rising concentration of media ownership may void this assumption. This observation motivates the study of the impact of ownership on the global media landscape and its influence on the coverage the actual viewer receives. To this end, the selection of reported events has been shown to be informative about the high-level structure of the news ecosystem. However, existing methods only provide a static view into an inherently dynamic system, providing underperforming statistical models and hindering our understanding of the media landscape as a whole. 

In this work, we present a dynamic embedding method that learns to capture the decision process of individual news sources in their selection of reported events while also enabling the systematic detection of large-scale transformations in the media landscape over prolonged periods of time. In an experiment covering over 580M real-world event mentions, we show our approach to outperform static embedding methods in predictive terms. We demonstrate the potential of the method for news monitoring applications and investigative journalism by shedding light on important changes in programming induced by mergers and acquisitions, policy changes, or network-wide content diffusion. These findings offer evidence of strong content convergence trends inside large broadcasting groups, influencing the news ecosystem in a time of increasing media ownership concentration.

\end{abstract}



%
%


\maketitle

\begin{figure}
\begin{minipage}{1.0\linewidth}
  \centering
  \includegraphics[width=\textwidth]{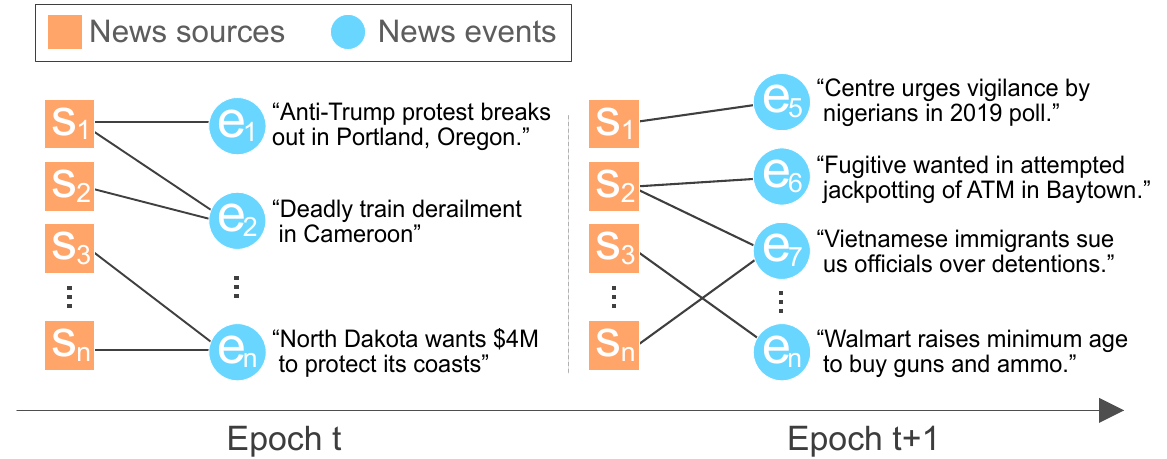}
  \caption{\textnormal{Illustration of the setting: we observe the coverage of news events from a fixed set of sources over several time epochs. Example events are extracted from the GDELT database.}}
  \label{fig:illustr}
\end{minipage}
\end{figure}

\section{Introduction}
\label{sec:intro}


Information about world events is reported as a continuous stream, processed then broadcast to an audience through a large number of channels and across various mediums. As no coverage could claim to be exhaustive, sources must first filter the stories that will be disseminated. The selection made by each channel is partial by nature. However, the variety of outlets, each with a wide array of considerations in the choice of its reporting, is assumed to ensure the diversity of news to which the reader is exposed to. This principle is often referred to as the \textit{external pluralism} assumption. It should ensure heterogeneity in the media space, encapsulating anything from the diversity of ownership to the independence of the editorial board.~\footnote{As opposed to \textit{internal pluralism}, where sources are assumed to present a wide variety of ideological viewpoints, communicated through different mediums.~\cite{mediaconcentration}} \textit{External pluralism} is also known as the ``supplier'' pluralism, since it should exclude the possibility of large broadcast groups exerting influence on downstream reporting. Yet in practice, this assumption does not always hold.
News channels are often owned or operated by commercial, private entities, implying that the ecosystem as a whole is influenced by economically motivated forces, such as mergers, acquisitions, or regulatory actions. The increase in concentration of ownership has been observed to be the dominating force in the media landscape, as reported by the Pew Research Institute in a 2017 study on the acquisition of local television stations.~\footnote{http://www.pewresearch.org/fact-tank/2017/05/11/buying-spree-brings-more-local-tv-stations-to-fewer-big-companies/}

While the literature, discussed in more detail in Section~\ref{sec:rel_work}, still debates the causal effects of market ownership structures on the diversity of offerings, the Federal Communications Commission (FCC)~\footnote{The FCC is the regulating body for multimedia communications in the United States.} has defended the idea that ``there is a positive correlation between viewpoints expressed and ownership of an outlet''.~\footnote{FCC, Biennial Media Ownership Order (2003)} This uncertainty strongly motivates the development of novel and interpretable methods, which would allow the observation of large-scale movements in the media landscape and allow correlating the observations with real-world factors.

Recent studies have tackled the problem of evaluating similarities across news sources~\cite{bourgeois2018selection, saez2013social} and have found that the analysis of coverage patterns can highlight non-obvious relationships between sources, such as the aforementioned corporate ownership structures. However, they have only captured a static snapshot of the media landscape. This hinders the ability to observe changes in content diffusion, and limits attempts to correlate the observations with potential factors in the news space. These models' predictive power is also limited by their inability to incorporate temporal dependencies. In other words, it restricts them to an informative but partial and static view of the news ecosystem. Indeed, the model of any dynamic system requires a time-dependent formulation, necessary to detect variations and measure trends. Existing approaches are blind to such transient effects.

Studying the effect of real-world factors on content diffusion is a challenging task. It requires the detection of subtle variations in the coverage patterns of individual channels, which can occur over long time spans. Any attempt to characterize these effects would suffer from the absence of a ground truth, prompting the need for a comparison across news channels. While similarities between channels are straightforward to compute for short, consistent periods of time, the continuous occurrence of new and unpredictable events breaks the ability to reason across successive snapshots of the news. A successful estimate of similarity over time should hence maintain temporal consistency of pairwise distances across news channels. At any given point in time, it should ideally do so without compromising its predictive accuracy compared to static methods. These challenges and constraints motivate the use of the models presented here.


\xhdr{In this work}, we propose a dynamic embedding model of the media landscape. By predicting news sources' coverage, the model captures their similarities throughout the observed period. The embedding space is maintained consistent over time by augmenting the model with a knowledge of previous time steps, and by adding a temporal regularization on the model's parameters. This improves the model's predictive capabilities and provides a temporally coherent source-wise similarity metric, allowing the visualization and analysis of long-ranging fluctuations in the news ecosystem. We also propose a systematic method to detect abrupt transition patterns in this similarity space. This enables the analysis of the news ecosystem beyond domain-specific knowledge and hand-crafted analysis.

In the context of news reporting, temporally-aware models provide a powerful tool for correlating content programming changes and real-world factors, from individual news outlet's behaviors to large-scale dynamics. Shining a light on these movements would naturally guide the search for source-level context, helping to characterize and identify the observation's driving forces. This information could then provide context to journalists investigating the coverage of a particular story or feed watchdog-like processes that monitor the health and evolution of the media landscape. To illustrate, we provide some prototypical questions which could arise in a journalistic probe, and that could be elucidated by information derived from the proposed model:


\begin{itemize}
\item What effect does the ownership of a news channel have on its content diffusion?
\item Which sources are most similar (resp. dissimilar) to a sample source, and how has this similarity evolved over time?
\item Which are the most consistent (or varying) news channels in terms of broadcast content?
\item Which broadcast groups exert a large influence on the content of their respective channels?
\end{itemize}

The rest of this paper is organized as follows. In Section~\ref{sec:rel_work}, we present relevant work from the literature. In Section~\ref{sec:data}, we describe the dataset used in our study. In Section~\ref{sec:model}, we introduce our model. In Section~\ref{sec:results}, we discuss the predictive performance of the model. In Section~\ref{sec:analysis}, we illustrate the use of our model in the identification of abrupt changes the media landscape, attempting to interpret these changes in coverage in Section~\ref{sec:discussion}. In Section~\ref{sec:conclusion}, we conclude and propose future research directions.
\vspace{-0.15cm}

\section{Related Work}
\label{sec:rel_work}
\xhdr{The study of the evolution of the media landscape} has led to prolific research lines spanning various fields. Early work had a tendency to examine the media ecosystem in isolation. For example Steiner's seminal work~\cite{RePEc:oup:qjecon:v:66:y:1952:i:2:p:194-223.} studies the interplay of consumer preferences and in-market competition on the diversity of radio broadcasting. This study omitted the role of external driving forces, which were only later modeled by Anderson \& Coate.~\cite{10.2307/3700696} They predicted that media consolidation, while economically beneficial for the market, would reduce competition and hence diversity for the viewer. This insight was later formulated in terms of ideological bias by Gentzkow \& Shapiro~\cite{RePEc:ecm:emetrp:v:78:y:2010:i:1:p:35-71} in the case of newspapers.

The tendency to integrate external driving factors has picked up steam in recent years, most notably with a theory of convergence in the media ecosystem. This convergence expresses itself through two seemingly contradictory features. On one hand, information is being delivered through an ever-increasing number of channels and means of diffusion. On the other hand, media ownership concentration has seen an upwards trend, with a large proportion of channels being owned by only a handful of media conglomerates.~\cite{pew2013} This dichotomy has been studied by, among others, Jenkins.~\cite{jenkins2004cultural} The author proposed a sketch of the phenomenon that looks further than the sole technological influence, reaching for larger cultural factors. Vizcarrondo et al.~\cite{vizcarrondo2013measuring} have more specifically investigated the concentration of media ownership. They reported on changes in the diversity of ownership within the media industry covering the 1976 through 2009 time period. 

\begin{table}[t]
\centering
\begin{tabular}{lp{0.7\linewidth}}
\toprule
Notation 			 & Description                 							\\ \midrule
$R$       			 & Interaction matrix $\in \mathbb{R}^{|S| \times |E|}$  \\
$S$ / $E$      		 & News source / Event set                    					\\
$E^+_{s_i}$          & Set of events covered by source $s_i$	\\
$s_i$     			 & Source $s_i \in S$ 									\\
$e_k$     			 & Event $e_k \in E$ 									\\
$K$       			 & Number of latent factors 								\\
$\hat{x}_{s_i, e_k}$ & Predicted preference of source $s_i$ for event $e_k$ 	\\
$D$ / $D_s$ 		 & Training / Testing set	\\
$\gamma$             & STD of the Gaussian random walk	\\
$\alpha$             & Learning rate	\\
$\lambda_{\Theta}$   & $\ell_2$-regularization weight	\\
$\lambda_{T}$        & Temporal regularization weight	\\
$\Theta$             & Set of model parameters	\\
\bottomrule
\end{tabular}
\caption{\textnormal{Notation} \label{notation-table}}
\vspace{-1.0cm}
\end{table}

A large body of work has also been introduced to study the effect of an ideologically slanted press. For instance in a large-scale observational study, DellaVigna et al.~\cite{DellaVigna2005TheFN} measured the effect of the introduction of a conservative-oriented channel (Fox News) led to gains of 0.4 to 0.6 percentage points in Republican voting in the towns where the channel was being broadcast.
While specific to a particular orientation, this work is in line with studies showing the profound influence of the media in voters' political awareness~\cite{mondak95} and their participation in the electoral process.~\cite{georgewald08, NBERw12317, NBERw15544}

The Federal Communications Commission (FCC) regularly issues studies regarding the state of the news ecosystem. Specifically, some of these studies focus on the effect of ownership on local news stations' content programming behaviors. However, by the authors' own admission~\footnote{``[a] larger number of independent owners will tend to generate a wider array of viewpoints in the media than would a comparatively smaller number of owners. We believe this proposition, even without the benefit of conclusive empirical evidence.'' FCC, Biennial Media Ownership Order (2003)}, these works often lack the breadth required by a large-scale empirical study. For example, Pritchard~\cite{Pritchard} conducted a study of the diversity of coverage for cross-owned media outlets during the 2000 presidential campaign but on a sample of only 10 newspapers. Groseclose \& Milyo~\cite{RePEc:oup:qjecon:v:120:y:2005:i:4:p:1191-1237.} proposed a measure of media bias which was evaluated on a set of 8 newspapers. Djankov et al.~\cite{Djankov03} did survey the news ecosystem on large scale, building a map of media ownership in 97 countries around the world, but this work dates back to 2003.

\xhdr{Dimensionality reduction methods} have been extensively used to model different perspectives or opinions. For example, Lahoti et al.~\cite{lahoti2018joint} rely on Non-Negative Matrix Factorization methods to learn a liberal-conservative ideology space on Twitter. In particular, the authors propose to approximate the users' ideology based on their online news consumption. 

The study performed by Saez-Trumper et al.~\cite{SezTrumper2013SocialMN} relies on a Principal Component Analysis (PCA) to detect similarities across news channels. In particular, they performed their analysis by covering three different types of biases in the press: \textit{gatekeeping bias}, \textit{coverage bias} and \textit{statement bias}. The authors' bias-capturing method does present similarities to our approach, but is formulated as an unsupervised approach and does not tackle temporal variability. Recently, Bourgeois et al.~\cite{bourgeois2018selection} proposed a supervised, embedding-based method to capture similarities across news sources based on their respective coverage. This model was also designed statically and is hence unable to accurately model temporal variations in news coverage. 

Learning to produce a personalized ranking from positive interactions only is a well-studied problem that has been investigated in the context of recommendations from implicit feedback~\cite{pan2008one} and as One-Class feedback~\cite{Hu2008CollaborativeFF} in the context of recommender systems.
\vspace{-0.1cm}

\xhdr{Temporally-aware methods} have received increasing attention and many previous models have now been adapted to the temporal setting. The \textit{dynamic embedding}, proposed by Rudolph et al.~\cite{rudolph2017dynamic} as a variation of traditional embedding methods, is generally aimed toward temporal consistency. The method is introduced in the context of word embeddings, which are used to characterize the evolution of English language. The model is built upon the initial \textit{exponential family embeddings} model.~\cite{rudolph2016exponential}

The field of personalization has many examples of temporally-aware models since human preferences tend to evolve over time. For example, influential work from Koren et al.~\cite{koren2009collaborative} models the changing nature of preference through a linear drifting term. Another approach relies on the use of Tensor Factorization (TF),~\cite{dunlavy2011temporal, acar2009link, xiong2010temporal} in which the extra dimension models temporal patterns in the data. We do not consider TF-based methods as valid candidate approaches since we focus on the problem of grounding representations over time by penalizing unnecessary differences between successive solutions of the model. The temporal modelling capabilities of TF-based methods would predict the evolution of sources and introduce additional temporal variations, consequently degrading interpretability.

He et al.~\cite{he2016ups} introduced a temporally-aware model of a recommender system in order to capture the evolution of fashion trends. Similar to Koren et al.~\cite{koren2009collaborative}, the authors also proposed the addition a drifting term to the model. 
The authors later proposed the use of a higher-order Markov chain that captures both short- and long-term dynamics~\cite{he2016vista}. Note that both models make use of Bayesian Personalized Ranking (BPR~\cite{Rendle2009BPRBP}) for their respective optimization procedure. In the context of networks, Yu et al.~\cite{yu2017temporally} proposed a temporal factorization for analyzing the evolution of network structures.

\xhdr{The media response to global events} has been studied, with many using the same data-source as our work. However, these approaches have focused mainly on content-level analysis, leveraging the multimodal capabilities of the GDELT dataset (please refer to Section~\ref{sec:data} for more detailed information about the dataset). These works have monitored the media response to events such as protests,~\cite{Qiao:2015:GMD:2859846.2860085} natural disasters~\cite{kwak2014first} or conflicts.~\cite{doi:10.1177/0022343313496803, Keertipati2014MultilevelAO, yonamine2013nuanced} This data-source was also used to get a world-wide view on coverage of global issues like climate change.~\cite{OlteanuCDA15} 
\vspace{0.7cm}


\xhdr{Research Questions: }
Given the work above, several research questions are of our interest and have remained unanswered:
\begin{description}
\item \textbf{RQ1:} How can we model the global media landscape over time?
\item \textbf{RQ2:} Is media consolidation highlighted by the resulting latent representation?
\item \textbf{RQ3:} How to systematically detect abrupt deviations in content diffusion at a source-level?\\
\end{description}


\section{Data}
\label{sec:data}

In this section, we present the selected data source and explicit our data collection process, providing general statistics about the resulting dataset. 
\vspace{-0.1cm}
\begin{table}[H]
\begin{tabular}{@{}ll@{}}
\toprule
                                      & GDelt Dataset       \\ \midrule
\multicolumn{1}{l|}{\# sources}       & 7\,278              \\
\multicolumn{1}{l|}{\# unique events} & 174M                \\
\multicolumn{1}{l|}{\# interactions}  & 588M                \\
\multicolumn{1}{l|}{time span}        & Feb 2015 - May 2018 \\ \bottomrule
\end{tabular}
\vspace{0.1cm}
\caption{\textnormal{Meta-data about the dataset (after preprocessing)}}\label{tab:stats}
\end{table}
\vspace{-1.0cm}

\subsection{Data Source}

Recently, several event collection databases have emerged on the Web, making accessible to the general public a global view of the daily world events. In this work, we rely on the \textit{Global Database of Events, Language, and Tone} (GDELT~\cite{Leetaru13gdelt:global}), a large database of annotated news events. GDELT~\footnote{https://www.gdeltproject.org/} was selected since it is publicly available and covers a reasonably large timespan and geography, offering a larger set of sources~\cite{an2016two} to study compared to alternatives such as EventRegistry.~\footnote{https://eventregistry.org/} 

For decades, event coding was performed manually. In the 1990s, the first automated systems started to gain traction in the academic community, with initiatives such as the KEDS system.~\cite{schrodt1994political} Its successor was proposed in the form of \textit{Text Analysis By Augmented Replacement Instructions} (TABARI), which is the engine that runs event coding for GDELT. This framework is designed to process large amounts of text to extract the presence of pairs of \textit{actors} and \textit{verbs}. To do so it matches elements from user-provided dictionaries, which contain a massive collection of event protagonists (i.e. actors) ranging from recognizable named entities (e.g. \textit{Barack Obama}) to functional placeholders (e.g. \textit{a local woman}). These actors are able to interact with the world through \textit{verbs} (i.e. actions), which can be self-contained (e.g. \textit{announces their intent to}) or involve a second actor (e.g. ~\textit{criticizes their opponent}). Several standards exist for these dictionaries. GDELT uses the \textit{Conflict and Mediation Event Observations} (CAMEO~\cite{Gerner02conflictand}).~\footnote{An exhaustive list of the considered categories can be found at http://data.gdeltproject.org/documentation/CAMEO.Manual.1.1b3.pdf} Note that, as a remnant of previous hand-curated event annotation frameworks~\cite{Laurance1990}, TABARI also provides an interface for manual hand-off to domain experts if the sentences become too complex. This reinforces GDELT's ability to uniquely annotate even the most fine-grained events.

GDELT also augments every news event it tags by extracting meta-information about the article including, but not limited to, its location, its tone, its Goldstein Scale~\cite{10.2307/174480} and refences the URL the event was scanned at. It scours a wide array of sources, from television stations to blogs, news wires and papers. Thanks to the information provided by this augmented event coding framework, GDELT assigns, for each news event, a global identifier, which makes it possible to link the same event's coverage across different news sources. Beyond the rich annotations provided by GDELT, this tracking is central to our study given that we only work at the coverage level, without considering the content itself.




\begin{figure}[H]
  \begin{minipage}[b,valign=b]{0.45\linewidth}
  \centering
  \includegraphics[width=\textwidth]{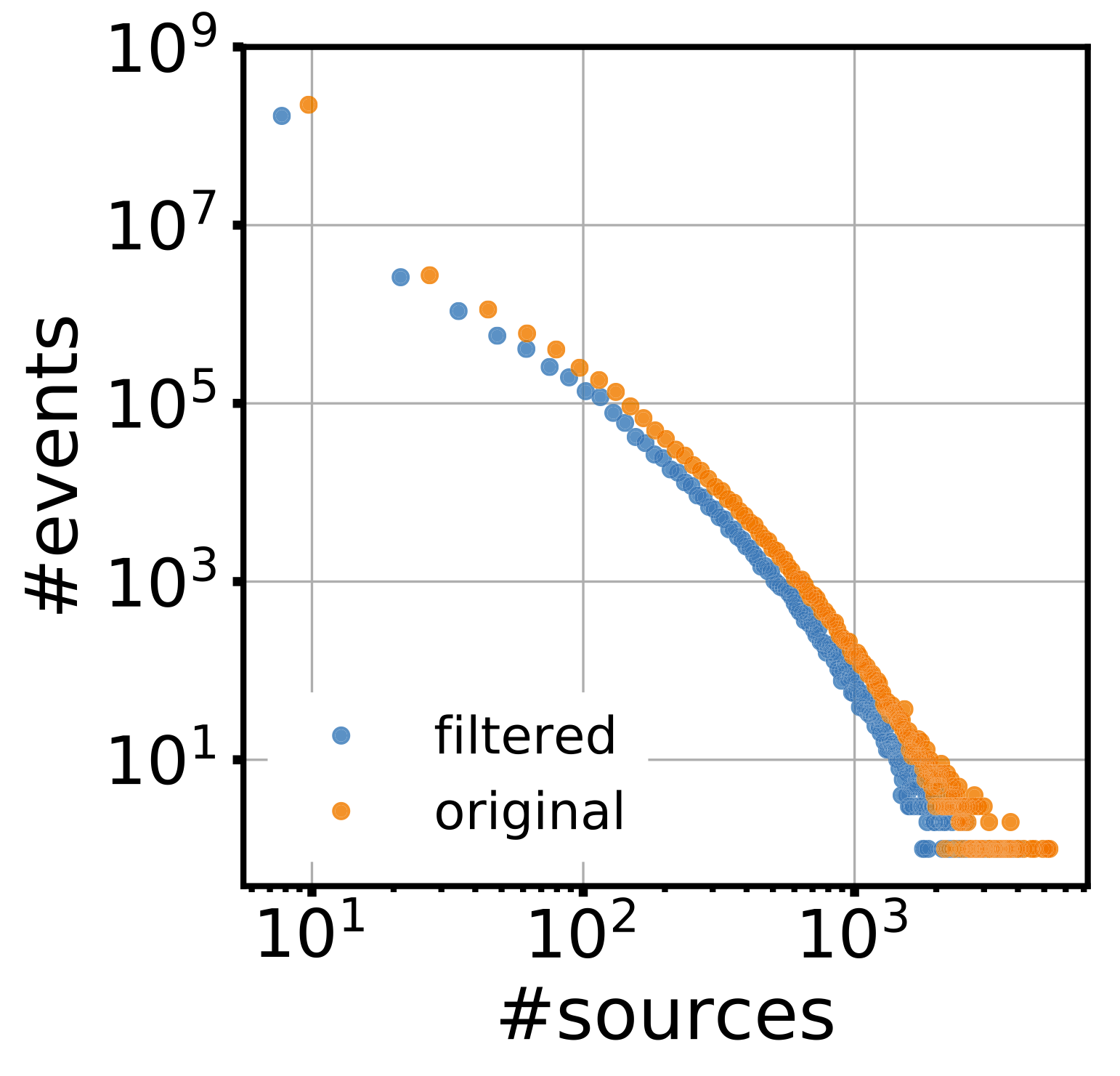}
  \caption{\textnormal{Number of sources covering an event.}}\label{fig:distrib_e}
\end{minipage}
\hfill%
\begin{minipage}[b,valign=b]{0.45\linewidth}
  \centering
  \includegraphics[width=\textwidth]{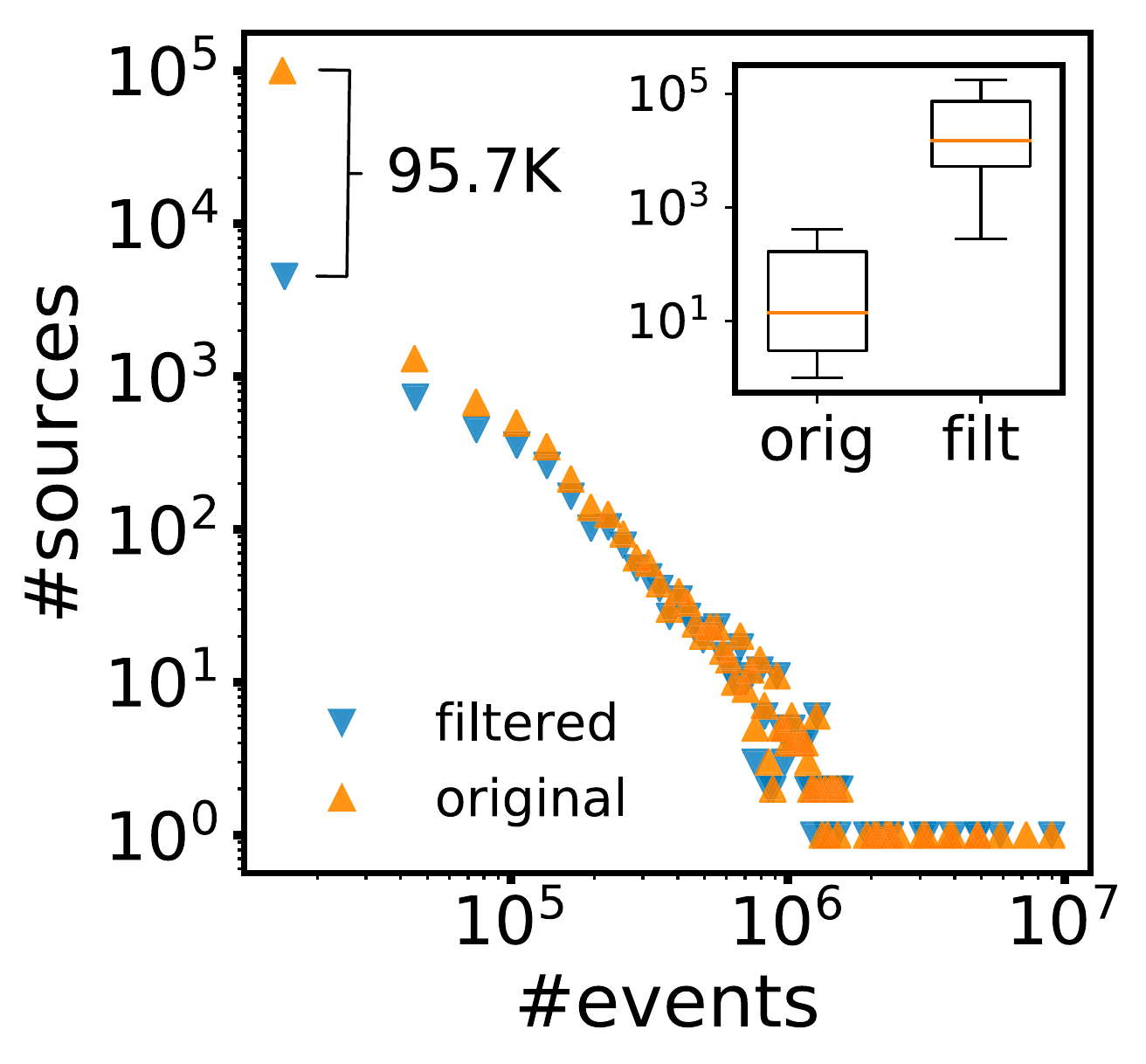}
  \caption{\textnormal{Number of events covered per source.}}\label{fig:distrib_s}
\end{minipage}
\end{figure}

\subsection{Data Preprocessing}

From the massive resource maintained by GDELT we can gather a dataset of interactions between sources and events, recording which sources covered which uniquely identifiable events. We focus our data collection campaign on the publicly available dumps of GDELT 2.0, released every 15 minutes since February 2015.  General statistics about the dataset are presented in Table~\ref{tab:stats} as well as Fig.~\ref{fig:distrib_e} and Fig.~\ref{fig:distrib_s}.

When considering the full time-span, GDELT references more than 105K different news sources. This represents a considerable increase compared to the 63K sources reported by Kwak et al.~\cite{an2016two} in 2016. However, as is shown in Fig.~\ref{fig:distrib_s}, most of the sources have only published a few articles over the relevant stretch. To maintain a consistent number of channels over time, we discard all channels inactive in any one of the time slices from our dataset. This retains around 7\,278 news sources in our dataset. The filtering step does remove a large fraction of available channels, but it mostly affects sources with a very low publishing rate: despite preserving only $7\%$ of the total channels, the selection still accounts for more than $76\%$ of interactions in the dataset (see Fig.~\ref{fig:distrib_s}). 

The dataset is split into slices with a duration of one month. This allows for a decent trade-off between having a significant amount of events covered in the training set, while also providing enough samples to observe time-dependent changes over the considered period. In principle, this scale could be modified to study the media landscape at different granularity levels. For example, a more fine-grained split might allow the observation of changes correlated with specific events. However, we choose to leave this analysis as future work given the significant amount of computation that the model requires.

\section{Model}
\label{sec:model}

In this section, we first motivate our approach. Then, we briefly introduce the use of Matrix Factorization methods in a ranking setting by formally describing the task and its objectives. Finally, we propose two strategies to extend the model to a temporal setting.

\subsection{Motivation}
The main focus of our work is to establish a dynamic model of coverage similarities across news channels in order (i) to uncover congruent coverage patterns in the pool of available channels, (ii) to establish a dynamic, predictive model of coverage and (iii) to provide a systematic methodology to identify and interpret structural changes in the media landscape. Recent political and societal focus on media accountability and transparency provide strong motivators in the development of such tools (see Section~\ref{sec:intro}). 

We model the interrelationships between sources and events by relying on a Matrix Factorization (MF) method. MF methods represent a natural way of projecting two disjoint sets of items in a common latent space of $K$ dimensions in order to model their interactions. Such personalized methods are commonly use by recommender systems, which routinely aim to model retail purchasing decisions. Used in our context, they model coverage decisions instead. We cast the problem to a One-Class learning setting~\cite{pan2008one} since we observe positive interactions only. The One-class formulation avoids making assumptions about negative examples: we do not distinguish between real negatives (i.e. the source purposely didn't cover an event) and unobserved interactions (i.e. the source wasn't aware of the event).


\subsection{Problem Statement}
Let us consider a set of news sources $S$ and a set of events $E$. Interactions between the two are represented by an interaction matrix $R \in \mathbb{R}^{|S|\times|E|}$. Observations take the form of dyadic interactions $(s_i,e_j), s_i \in S, e_j \in  E$ which express source $s_i$'s coverage of event $e_j$. Equivalently, we define in matrix form that $R_{i,j}=1$ if source $s_i$ has covered event $e_j$ and $R_{i,j}=0$ otherwise. 

Predicting the unobserved entries of matrix $R$ is achieved by taking the dot-product of two low rank matrices, such that \linebreak $ R \approx P^T \cdot Q$, where $P \in \mathbb{R}^{K \times |S|}$ and $Q \in \mathbb{R}^{K \times |E|}$ with $K << |S|,|E|$. Every source $s_i$ (resp. every event $e_j$) is represented by a column in $P$ (resp. $Q$). We will refer to these columns as an embedding vector throughout the remainder of this work. We will refer to $\Theta$ as the set of parameters for our MF predictor, such that $\Theta = \{P,Q\}$.

\xhdr{Objectives: } The model is trained with the objective of predicting the likelihood of a source covering a particular event. The predicted likelihood $\hat{x}_{s_i, e_j}$ of source $s_i$ covering event $e_j$ is computed as the dot-product between the two respective embedding vectors,

\begin{equation}
\hat{x}_{s_i, e_j} = p_{s_i}^T \cdot q_{e_j}\,.
\end{equation}

Instead of best approximating the reconstruction of matrix $R$, this objective is stated as a ranking problem in which positive examples should obtain a higher rank than negative ones, i.e. to predict a higher score for an event that has been covered than for random negative samples. Optimizing a MF model with a ranking criterion is equivalent to maximizing the following probability,

\begin{equation}
\begin{split}
\Pr(e_j>_{s_i}e_k|\Theta) \, & \coloneqq \, H(\hat{x}_{s_i,e_j} - \hat{x}_{s_i,e_k}) \\
 & \; \approx \,\, \sigma(\hat{x}_{s_i,e_j} - \hat{x}_{s_i,e_k}) 
\end{split}
\end{equation}

where $e_j$ is an event covered by source $s_i$ and $e_k$ is a randomly sampled negative event; formally, $e_j \in E^+_{s_i}$ and $e_k \in E \setminus E^+_{s_i}$. We adopt the notation $>_{s_i}$ to denote a source $s_i$ preferring to cover $e_j$ over $e_k$ and model the observation of this preference using $H(\cdot)$, the Heaviside step function: $H(\cdot)$ is equal to $\mathbf{1.0}$ for positive inputs and to $\mathbf{0.0}$ otherwise. Therefore, $H(\hat{x}_{s_i,e_j} - \hat{x}_{s_i,e_k})$ would always be equal to $\mathbf{1.0}$ for an ideal predictor. In practice, $H(\cdot)$ is approximated by the differentiable logistic sigmoid $\sigma(\cdot)$.



Finally, we maximize BPR, our log-likelihood criterion

\begin{equation} 
\text{BPR} \, \coloneqq \, \sum_{(s_i,e_j,e_k) \in D} \ln \sigma(\hat{x}_{s_i,e_j} - \hat{x}_{s_i,e_k}) - \lambda_\Theta \; ||\Theta||^2 \,.
\end{equation}

Note the inclusion of an  $\ell_2$-regularization term over the set of parameters $\Theta$. We please refer the reader to the work of Rendle et al.~\cite{Rendle2009BPRBP} for more details about this optimization scheme.





\subsection{Temporal Setting}
In this section, we describe the adoption of the dynamic embedding scheme proposed by Rudolph et al.~\cite{rudolph2017dynamic} in the context of news coverage modeling. In particular, we adopt two strategies to maintain temporal consistency across time slices, respectively based (i) on a Gaussian random walk (\textit{RW}) and (ii) on the addition of a temporal regularization term (\textit{RG}). We adopt the notation $p_{s_i}^{(t)}$ to denote the embedding vector of news source $s_i$ at the $t$-th time step.

\xhdr{Prior on the embedding vectors: } Without information about former time slices, existing methods typically initialize embedding vectors to small, randomly distributed values. However, such approaches do not take advantage of any prior knowledge acquired during anterior training steps. The addition of a prior on embedding vectors represents a simple, yet powerful strategy to leverage previously acquired knowledge about sources. In particular, embedding vectors at the $t$-th time step are initialized using a Gaussian random walk around their final values at time step $(t-1)$. The Gaussian random walk is expressed as follows: 

\begin{equation}
p_{s_i}^{(t)} \sim \begin{cases}
    \mathcal{N}(0, \gamma^{-1} I), & \text{if t=0}\,.\\
    \mathcal{N}(p_{s_i}^{(t-1)}, \gamma^{-1} I), & \text{otherwise}\,.
\end{cases}
\end{equation}\\

This initialization scheme ensures a smooth transition of the parameter set learned in two consecutive time slices. This yields a more stable embedding space, offering a coherent expression of divergence across time-steps. Since events are inherently much more volatile than sources, we initialized their embedding vectors at random at each new time slice.

\xhdr{Optimization using temporal regularization: } 
The second part of the dynamic scheme takes the form of a temporal regularization term. The newly introduced term penalizes large variations across time steps by minimizing the distance of an embedding vector at the $t$-th step to its final value at step $(t-1)$. 

The final log-likelihood criterion, BPR-T, can then be formulated as follows for the $t$-th time split 

\begin{equation} 
\label{loss}
\begin{split}
\text{BPR-T}^{\;(t)} \coloneqq \sum_{(s_i,e_j,e_k) \in D} & \ln \sigma(\hat{x}_{s_i,e_j} - \hat{x}_{s_i,e_k}) \\
 & - \lambda_\Theta \; ||\Theta||^2 \\
 & - \underbrace{\lambda_T \| p_{s_i}^{(t)} - p_{s_i}^{(t-1)} \|^2}_{\text{temporal regularization}}
\end{split}
\end{equation}

\clearpage
The model is optimized using stochastic gradient ascent and is fitted once for every time split. Update steps are defined as follows:



\begin{equation} 
\label{update_step}
\begin{split}
q_{e_j}^{\;(t)} & \leftarrow q_{e_j}^{(t)} + \alpha (\sigma(-\hat{x}_{s_i,e_j,e_k})\cdot   \;\;\; p_{s_i}^{(t)} \, \;\, - \lambda_\Theta \; q_{e_j}^{(t)}) \\
q_{e_k}^{\;(t)} & \leftarrow q_{e_k}^{(t)} + \alpha (\sigma(-\hat{x}_{s_i,e_j,e_k})\cdot  (-p_{s_i}^{(t)}) - \lambda_\Theta \; q_{e_k}^{(t)}) \\
p_{s_i}^{\;(t)} & \leftarrow p_{s_i}^{(t)} + \alpha (\sigma(-\hat{x}_{s_i,e_j,e_k})\cdot (q_{e_j}^{(t)} - q_{e_k}^{(t)}) \\
 &  \qquad - \lambda_\Theta \; p_{s_i}^{(t)} \\
 &  \qquad - \lambda_T ( p_{s_i}^{(t)} - p_{s_i}^{(t-1)} ) ) \\
\end{split}
\end{equation}

where $\alpha$ is our learning rate. We use the notation $\hat{x}_{s_i,e_j,e_k}$ to denote the quantity $(\hat{x}_{s_i,e_j} - \hat{x}_{s_i,e_k})$. Note that triplets $(s_i,e_j,e_k), e_j \in E^+_{s_i}$ and $e_k \in E \setminus E^+_{s_i}$ forming the training dataset $D$ are randomly sampled during the optimization.~\footnote{We sampled both positive and negative examples uniformly. More complex sampling approaches exist \cite{rendle2014improving} but are outside of the scope of this work.}

\subsection{Evaluation}
To assess the performance of the different methods, we adopt a \textit{leave-one-out} methodology, in which a single event per source is withheld at random from the training set $D$ to constitute the test set $D_s$. This approach ensures that all sources have similar weights in the evaluation. We adopt the widely used \textit{Area Under the Curve} (AUC) as a measure of performance. In the context of this work, the evaluation procedure is formally defined as follows

\begin{equation} 
\label{auc}
\text{AUC} = \frac{1}{\left\vert{D_s}\right\vert} \sum_{\!\!\!\!\!\!\!\!\!\!\!\!\mathrlap{(s_i, e_j, e_k)\in D_s}} H( \hat{x}_{s_i,e_j} - \hat{x}_{s_i,e_k})\,.
\end{equation}

where $e_j$ is an event covered by $s_i$ and $e_k$ is an event that $s_i$ hasn't covered, randomly sampled at testing time. Negative samples are drawn uniformly at random across all unique event of the current time slice (we omitted the time indices for the sake of brevity).

\subsection{Experimental Setting}
\label{sec:experiment}
The code for experiment and analysis will be made available at publication time under an open-source license. All experiment-related code was run on a 6-core machine, equipped with an \texttt{Intel(R) Xeon(R) CPU E5-2630 @ 2.30GHz}, for a total training time of approximately 3 days.~\footnote{The training time is reported for the full 3-year period; a production-ready application would typically be optimized and use incremental updates instead.} We restricted the tuning of hyper-parameter $\lambda_T$ to a subset of values $\in \{0.001, 0.01, 0.1, 1.0\}$. All scores and figures are reported using $\lambda_T = 0.1$, which we found to provide the highest cross-validated accuracy. For computational reasons, the other parameters were coarse-tuned on a static snapshot and were set to $K=20$ and $\lambda_{\Theta}=0.1$, $\gamma=0.01$ and $\alpha=0.1$.

\begin{figure}[h]
\centering
    \begin{minipage}[b,valign=b]{0.45\linewidth}%
      \centering
		\includegraphics[width=\textwidth]{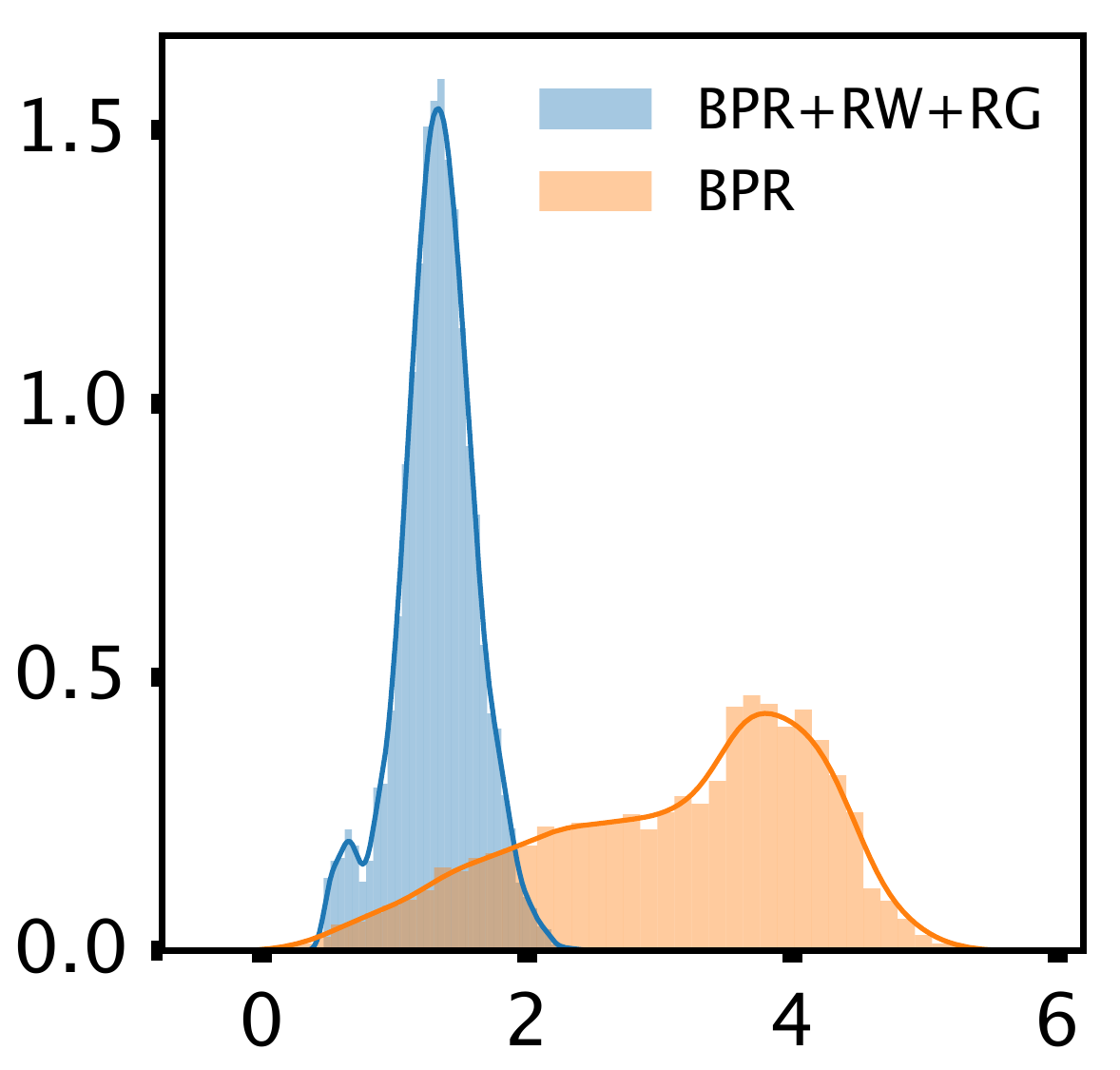}
        \caption{\textnormal{Average displacement of embeddings}}\label{fig:avg_disp}
    \end{minipage}%
  \hfill%
  \begin{minipage}[b,valign=b]{0.45\linewidth}%
    \begin{table}[H]
      \vspace{0.25in}
      \begin{tabular}{@{}ll@{}}
        \toprule
                                           & AUC             \\ \midrule
        \multicolumn{1}{l|}{POP}           & 0.6509          \\
        \multicolumn{1}{l|}{BPR}           & 0.8959          \\
        \multicolumn{1}{l|}{BPR + RG}      & 0.9089          \\
        \multicolumn{1}{l|}{BPR + RW}      & 0.9318          \\
        \multicolumn{1}{l|}{BPR + RW + RG} & \textbf{0.9337} \\ \bottomrule
        \vspace{0.15in}
      \end{tabular}
      \caption{\textnormal{Method contribution to performance}}\label{tab:scores}
    \end{table}%
    
  \end{minipage}%
\end{figure}

\begin{figure}[b]
	\includegraphics[width=0.45\textwidth]{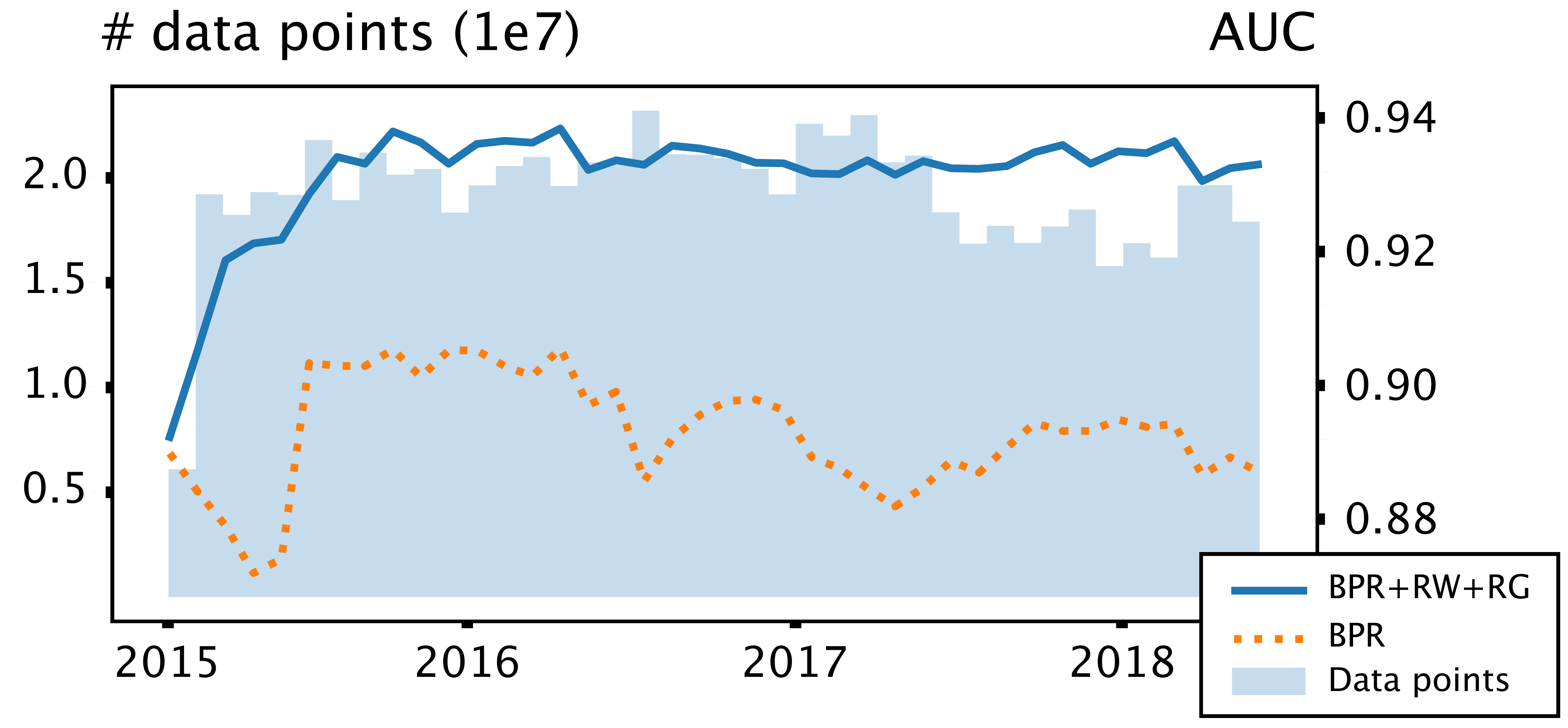}
	\caption{\textnormal{Performances (AUC) and dataset size per month}}\label{fig:auc_time}
\end{figure}

\begin{figure*}
\centering
\includegraphics[width=\textwidth]{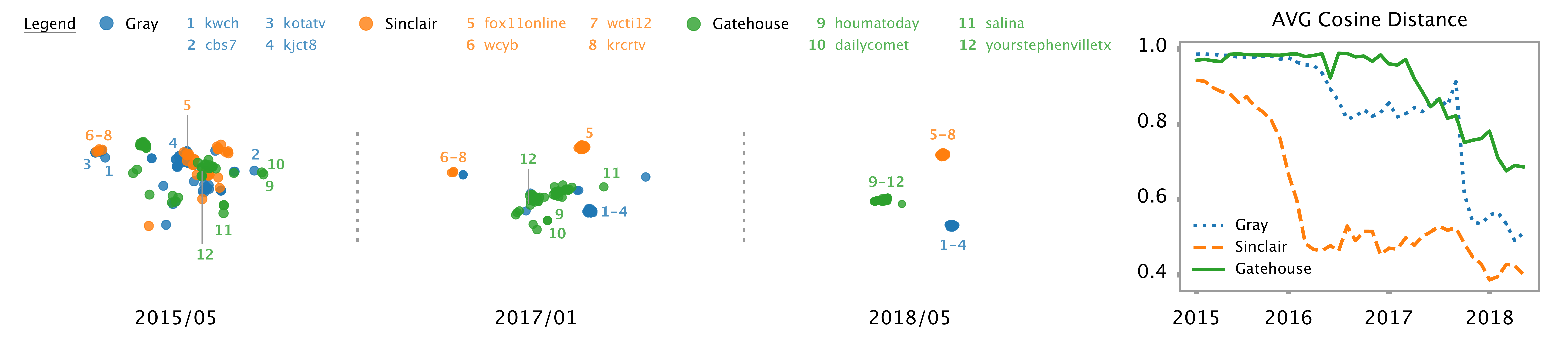}
\hfill%
\centering
\caption{\textnormal{Case-study of the influence of ownership on the selection of covered events (best seen in color)} 
\textnormal{ \newline\textbf{Left:} Sample of the temporal evolution of the media landscape, learned with dynamic embeddings, illustrated with t-SNE.~\cite{tsne} To test the effect of ownership on diversity of coverage, we monitor the evolution of a set of sources all owned by the same broadcast group in 2018, backtracking their evolution in embedding space. Starting with an initial seed source known to be operated by - or affiliated to - one of three large American media conglomerates (Gray Television Inc., Sinclair Broadcasting Group and GateHouse Media), we build a set of 50 sources per group so as to have a comparably sized sample for each. These sources are also all verified to be owned - or operated by - one of these groups at the last observed time-frame of our dataset (May 2018) by cross-checking publicly available information. We project their positions from the embedding space onto a display-friendly plane, showcasing snapshots of their movement over time, as they collapse into highly similar and cohesive clusters in the similarity space.} \newline \textnormal{\textbf{Right:} Average inter-source cosine distance between sources of each group in the dynamic embedding space, over time.}}
\label{fig:flipbook}
\end{figure*}



\section{Results}
\label{sec:results}
In this section, we compare the existing static embedding model with the dynamic method presented in Section~\ref{sec:model}. The effect of both the prior on the embedding vectors and the temporal regularization are then measured in isolation.

Table~\ref{tab:scores} summarizes the performances of the various approaches by taking the mean AUC scores obtained, for each month, over the considered period. \textit{BPR} denotes the core of the algorithm without any temporal component. \textit{RG} denotes the use of temporal regularization and \textit{RW} denotes the use of a Gaussian random walk for embedding initialization. To compare with a non-parametrized approach, we also include \textit{POP}, a popularity-based baseline: the score for a given event is a function of its frequency in the training set.

Several observations can be made in light of these results. Firstly, the combination of the two strategies \textit{RG} and \textit{RW} is shown to provide the best observed predictive performances. Secondly, the individual strategies do not provide the same performance improvements. Results suggest that a proper initialization contributes much more to an accurate prediction than strong temporal regularization. In parallel experiments, we even observed that an increase of $\lambda_{T}$ decreases performances. We hypothesize that this is due to the model's inability to handle abrupt changes in source behavior, since a strong regularizer would penalize a large difference with respect to the previous time step.
\vspace{0.2cm}

The use of the proposed dynamic strategies (\textit{RG} and \textit{RW}) provides better overall consistency of the latent space across time slices. This is visible by measuring the average displacement of sources in embedding space. As shown in Fig.~\ref{fig:avg_disp}, sources are much more stable in the dynamic setting compared to the static embedding procedure. The added stability of the embedding space provides a usable expression of divergence across time-steps. This means source similarity can be coherently compared across the entire observed period, while also providing an overall improvement of the AUC scores (see Fig.~\ref{fig:auc_time}).

\section{Analysis}
\label{sec:analysis}


In the following section, we discuss the interpretation of the model introduced in Section~\ref{sec:model}. We first describe an approach to visualize the evolution of the news ecosystem. Then, we propose a systematic, unsupervised way to detect abrupt deviations in this space.
\vspace{0.2cm}

\subsection{Visualizing the Media Landscape}
\vspace{0.2cm}

We start by introducing an example case-study, which should illustrate the usefulness of the presented model in facilitating the understanding of the news ecosystem. The study is centered on three representative media conglomerates that we track throughout the 40-month period covered by our dataset: \textit{Gray Television Inc.} (over 100 television stations), \textit{Sinclair Broadcasting Group} (over 190 television stations) and \textit{GateHouse Media} (over 140 newspapers).~\footnote{as of August 2018.}

As a starting point, the model described in Section~\ref{sec:model} is optimized in its most successful setting (\textit{RW+RG}). Dimensionality reduction is performed in order to have a more interpretable view into the embedding vectors. In Fig.~\ref{fig:flipbook} - (\textit{left}) we use a t-distributed Stochastic Neighbor Embedding (t-SNE~\cite{tsne}), a popular method for the visualization of high-dimensional data. In order to maintain consistency across time steps in the projection, we initialize the parameters of the t-SNE optimization procedure at time slice $t$ with the final parameters of time step $t-1$. This seeds the projected points' positions instead of assigning them to a random initial position, allowing for easier tracking between time steps. 

Additionally, in order to avoid interpreting from model parameters only, which might be misleading due to optimization artifacts, we correlate sources' trajectories with the average pairwise cosine distance between sources of each group. These distances are computed using sources' respective sets of covered events for each month. Overall, this procedure allows us to coherently visualize the evolution of the media landscape over time, uncovering non-obvious dynamics at several scales, from the ecosystem as whole (e.g. convergence phenomenons) down to individual sources (e.g. shift toward a group).

The designed visualization of the media landscape is presented in Fig.~\ref{fig:flipbook}. Please refer to Section~\ref{sec:discussion} for a more detailed discussion.
 

\begin{figure*}
\hfill%
\begin{minipage}[c]{1.0\linewidth}
\includegraphics[width=\linewidth]{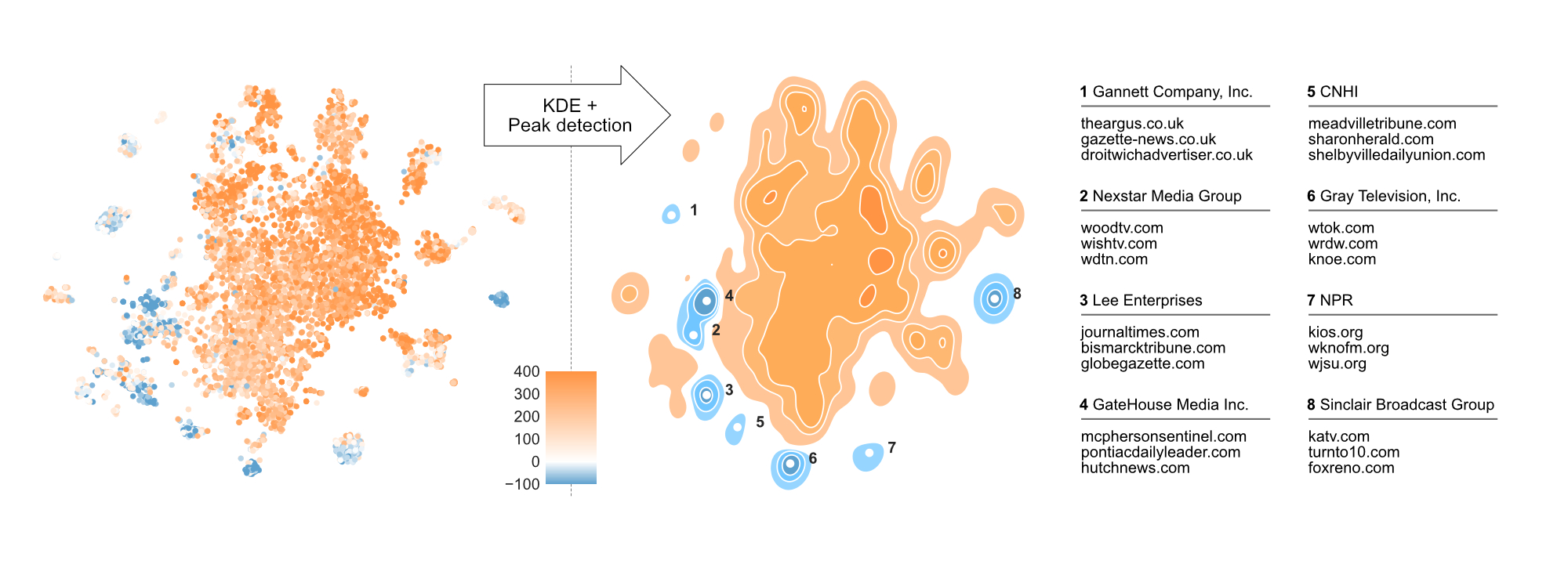}
\hfill%
\end{minipage}
\vspace{-0.8cm}
\caption{\textnormal{\textbf{Left:} t-SNE~\cite{tsne} projection of the embedding space for May 2018, colors represent the weight $w_{s_i}$ of each source $s_i$. \textbf{Center:} Identified attractor map built from the attraction potential map ($\sigma=1.9$, $k=500$). \textbf{Right:} Detail and affiliation of attractors identified in (\textit{center}), with the set of 3 sources closest to the uncovered poles. Best seen in color.}}
\label{fig:scatter_density}
\end{figure*}

\subsection{\large{Detecting Fluctuations in the Media Landscape}}
\label{ssec:fluctuations}
\vspace{0.2cm}

Even with extensive domain knowledge, tracking the evolution of a group of sources belonging to specific entities remains a tedious task, since it requires the manual identification of sources of interest and the validation of their common factors. Therefore, in the following section, we propose an unsupervised method which leverages the models' a priori knowledge to identify abrupt changes in sources' content diffusion patterns. In particular, the proposed framework aims to identify attractors, e.g. sources that tend of attract others in latent space, suggesting an alignment of coverage.

\xhdr{Attractors: } News channels involved in a consolidation of resources typically tend to have increasingly similar coverage patterns. As seen in Fig.~\ref{fig:flipbook} - (\textit{right}), the phenomenon manifests itself as the convergence of a subset of sources toward a common position in embedding space. Systematically detecting such gatherings around a common location, that we will loosely refer to as attractors, would allow to interpret each of these patterns in isolation. We propose a method in two steps. 
Firstly, we identify sources whose distances to other channels are abruptly reduced at any point in time. Secondly, we identify the absolute position towards which those sources tend to converge. \\\\



We first define the matrix $Z \in \mathbb{R}^{|S|\times|S|}$ that represents the cumulative difference of distances between any two news sources over time. We compute such distances in relative terms in order to avoid any drift component from the evolving latent space. More formally, we define $Z$ as follows.

\begin{equation} 
Z_{i,j} = \sum_t \quad \lVert p_{s_i}^{(t)} - p_{s_j}^{(t)} \rVert - \lVert p_{s_i}^{(t-1)} - p_{s_j}^{(t-1)} \rVert\,,
\end{equation}



where $\lVert \cdot \rVert$ is the Euclidean norm. In this setting, two sources whose distances are consistently reduced would produce a negative value of large magnitude in their corresponding entries of $Z$. Therefore, we rely on the matrix $Z$ to identify channels having undergone large reductions of distance with other sources. In more details, we retrieve from $Z$ the $k$-sources having the largest negative cumulative difference with any other source, i.e. the minimal value of each row in $Z_{i,j}$. By taking into consideration only the top-$k$, we can capture large shifts only and avoid considering small movements due to random factors.
Once identified, the relative displacement of these $k$ sources can be visualized in latent space. In particular, each source $s_i$ in our dataset will be qualified by a single weight $w_{s_i}$, computed as the sum of the cumulative difference in $Z$ with respect to the $k$ considered sources. As shown in Fig.~\ref{fig:scatter_density} - (\textit{left}), negative values reveal sources that tend to exhibit agglomerative behaviors.
\vspace{0.2cm}

Until now, we observed strong fluctuations in terms of inter-source distances. The next step is to define a systematic way of identifying the centers around which these shifts occur, in absolute terms and at any given point in time. On a 2D projection of sources, we apply a weighted Kernel Density Estimator (KDE), a non-parametric method for estimating the Probability Density Function from a set of samples, under weak smoothness assumptions. The objective is to detect areas containing a high density of sources with negative weights and locate their density peaks. We use the weights $w_{s_i}$ as input of the estimator. The bandwidth selection is a function of the data's covariance, multiplied by a constant factor $\sigma$.~\cite{Silverman86} An example of the resulting density is presented in Fig.~\ref{fig:scatter_density} - (\textit{center}). Finally, local extrema are collected using a local minimum filter, a simple method routinely used in computer vision. The set of identified poles and the top-$3$ closest sources surrounding them is shown in Fig.~\ref{fig:scatter_density} - (\textit{right}).
\vspace{0.2cm}

\begin{figure*}
\begin{minipage}[c]{0.32\linewidth}
    \includegraphics[width=\linewidth]{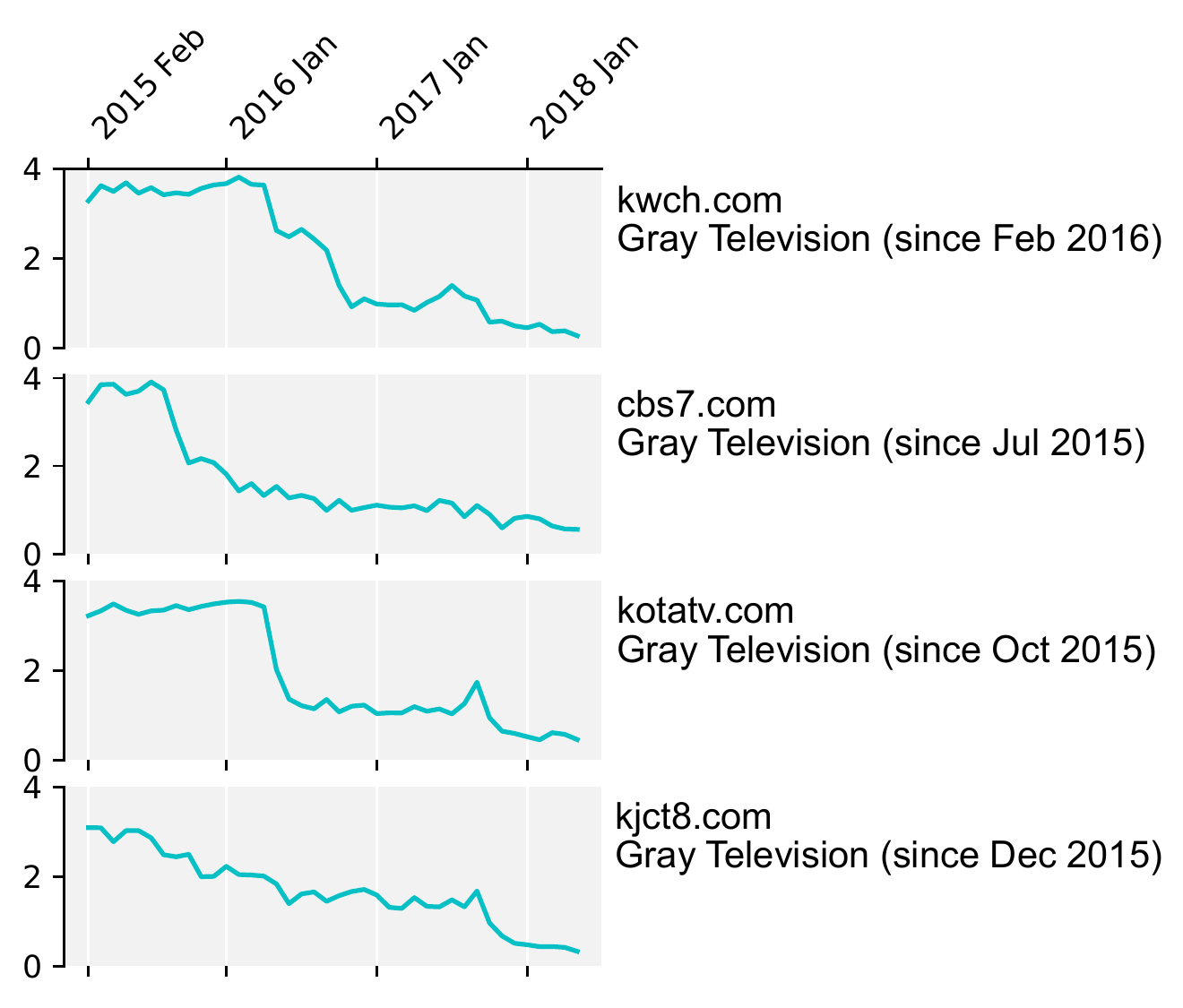}
\end{minipage}
\hfill%
\begin{minipage}[c]{0.32\linewidth}
    \includegraphics[width=\linewidth]{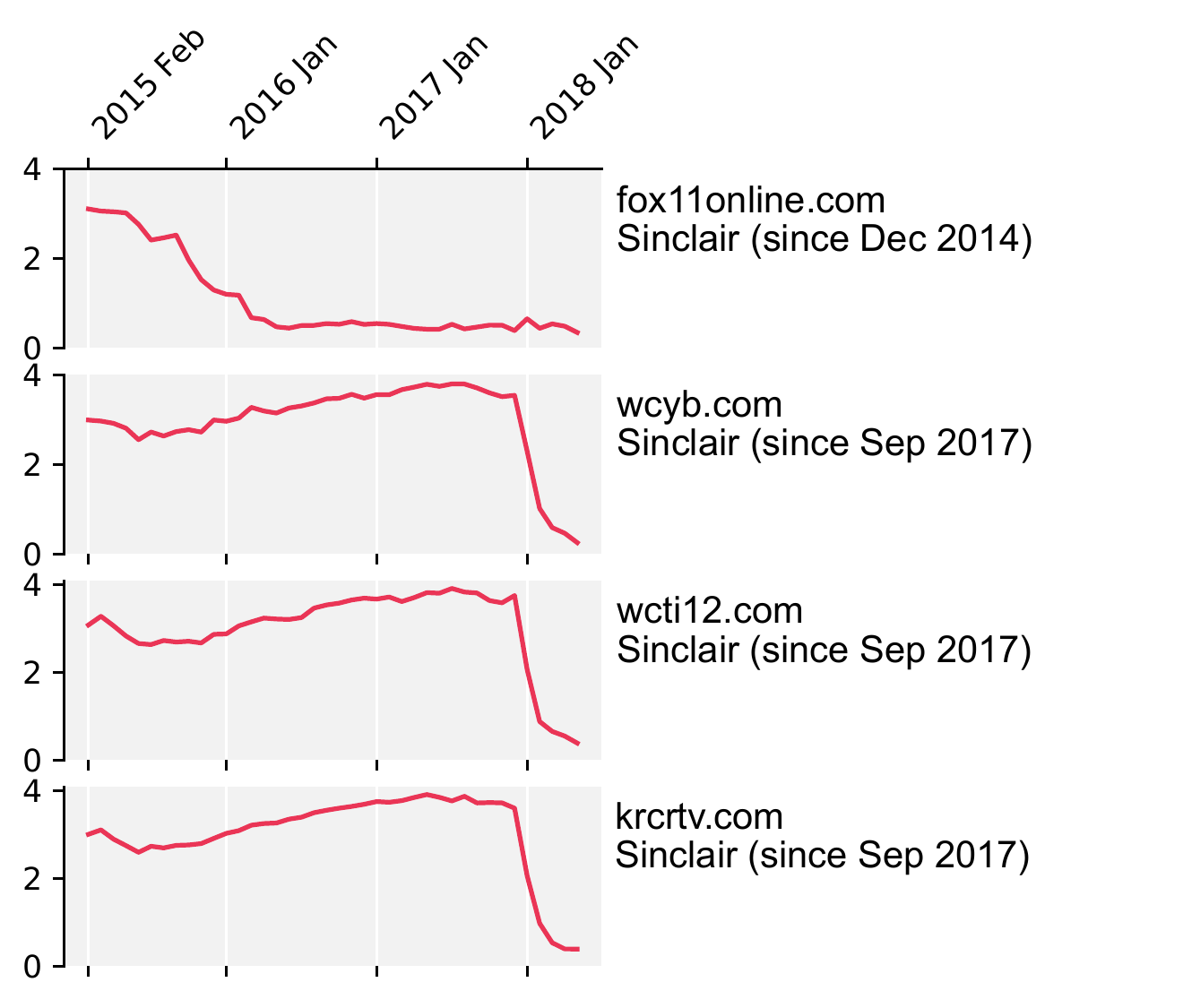}
\end{minipage}
\hfill%
\begin{minipage}[c]{0.32\linewidth}
    \includegraphics[width=\linewidth]{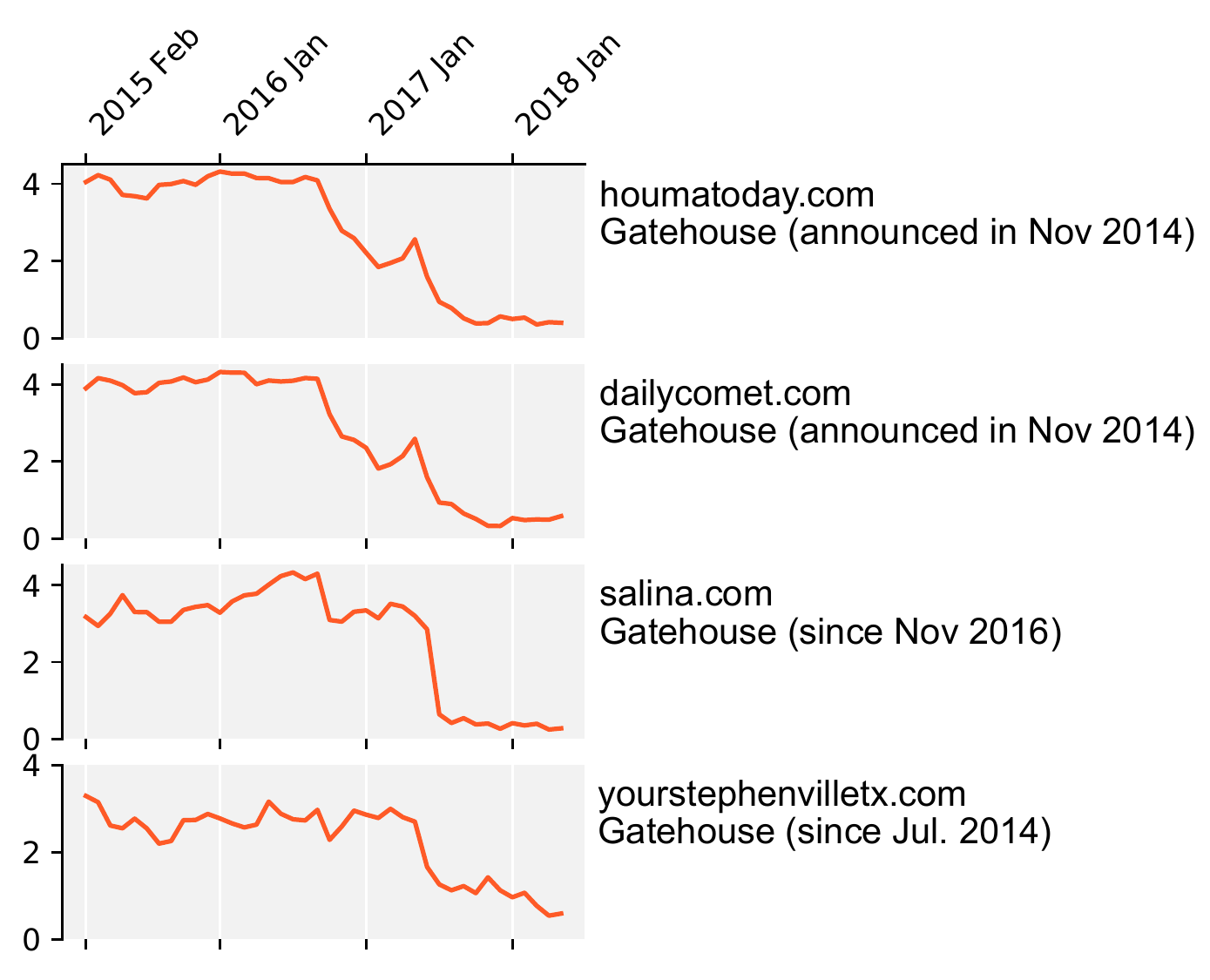}
\end{minipage}%
\caption{\textnormal{Euclidean distance from channels to their respective affiliations, in latent space, over time. Affiliation positions are computed as the centroid of 3 seed sources (taken from Figure~\ref{fig:scatter_density} - (\textit{right})). Those patterns are extracted without any supervision: retrieved channels are the ones having undertaken the largest shift of content over the considered period (2015-2018). Channels acquisition dates are given for reference.}}\label{fig:shift}
\end{figure*}

\xhdr{Attractees: } Having identified a set of attractors in latent space, the dual observation can be made: the identification of sources that experienced large movements in latent space toward any of the previously identified attraction poles. These are sources that have been strongly influenced by external forces, for example in the content consolidation phase after an acquisition. The detection of these phenomenon is done relative to a specific pole of attraction. In order to track the distance to a pole over time, we start with a set of seed sources. An obvious choice is to study the top-3 closest sources from the poles, detailed in Fig.~\ref{fig:scatter_density} - (\textit{right}). 
The ranking of sources having undergone a large shift can once again be made systematic. In particular, we rank sources according to the largest difference in distance to the pole between two consecutive time steps. The distance to the centroid of these sources yields the distance maps shown in Fig.~\ref{fig:shift} for the top-$4$ sources with the largest shifts. 

\vspace{-0.1cm}
\section{Discussion}
\label{sec:discussion}





The structure of the news landscape is in a constant state of flux. It is often difficult to follow the evolution of its organizational structure and even more so to determine what influenced these changes. In this section, we discuss how the fluctuations of broadcast patterns can be informative about channels' organizational structure. 
We report that important changes in this structure are identifiable through abrupt shifts in content diffusion, and showcase the models' ability to systematically highlight this variability in the coverage space.\\\\


\subsection{Observing the effects of ownership on the media landscape }
\vspace{0.2cm}

The selection and diffusion of events by individual news channels could be influenced by a large number of factors, from obfuscated economic drivers to convoluted distribution schemes. 
Theoretically, the external pluralism assumption would prevent large-scale organizational changes in news outlets from inducing significant shifts in coverage. Our findings question the validity of this assumption. \\

We provide evidence that ownership can indeed exert its influence on the content being distributed downstream. The most distinct and recurring pattern pointing to this conclusion is the subsequent alignment of coverage patterns after an outlet's acquisition. Some examples are clearly observable in Fig.~\ref{fig:flipbook}, such as the acquisition of 14 stations from the Bonten Media Group by Sinclair Broadcasting Group (SBG) in a deal completed on September 1st, 2017.~\footnote{https://tvnewscheck.com/article/103465/sinclair-buying-bonten-stations-for-240m/} Visible through the lens of a decrease in the average inter-source distance of the Sinclair stations, this consolidation of coverage can also be tracked in the embedding spaces' visualization in Fig.~\ref{fig:flipbook} - (\textit{left}). This can be observed in Fig.~\ref{fig:shift} as well, albeit with a slight delay with a sharp decrease in the distance from channels like \texttt{wcyb.com}, \texttt{wcti12.com} or \texttt{krcrtv.com} to the center of the Sinclair attraction pole (\#8 in Fig.~\ref{fig:scatter_density}).\\

We observe similar behaviors during others large-scale acquisitions, for example in the purchase of a group of assets from the Morris Publishing Group by GateHouse in August 2017.~\footnote{https://www.poynter.org/news/gatehouse-acquires-morris-publishings-11-daily-newspapers} Other observations of this phenomenon include the sudden increase in coverage similarity of Gatehouse-owned stations around April 2017 (see Fig.~\ref{fig:shift} - \textit{right}). While not directly correlated to a specific merger or acquisition, these movements could hint at a company-wide content alignment campaign.\\

Such observations also support the convergence hypothesis. The visualization in Fig.~\ref{fig:flipbook} exemplifies this effect: many of the sources that are present in one of these group's media portfolios start out from vastly different regions in embedding space. This is visible in their high initial average cosine distance in Fig.~\ref{fig:flipbook} - (\textit{right}) and their dispersed placement in Fig.~\ref{fig:flipbook} - (\textit{left}). In the last frame however, these same sources form highly coherent, tight groups in embedding space - and in the visualization. Despite the fact that this case-study back-tracks the evolution of sources across time, explaining the density of the last frame, their convergence points to a unification of coverage patterns over time.

\subsection{Detecting highly influential broadcast groups  }
\vspace{0.2cm}

News outlets present complex content distribution schemes, as is particularly visible in television broadcasting: the on-air content is produced by a wide range of affiliates, from well-known household names to in-house teams,~\footnote{A study conducted by Pew Research in 2014 already demonstrated a steady decline in locally produced content, with 1 in 4 local news stations not producing any of their own content~\cite{pew2013}.} being distributed through channels with another, often different set of owners. 
While the consolidation of broadcast material for economies of scale or investigative resources for economies of scope can be economically beneficial for the broadcaster, it is also potentially deceitful for a news consumer, as the exact origin of the broadcast content is not always known. By extension, the unique slant it carries in its selection of news is not clearly obvious. Not only does it carry is unique biases in terms of the way in which it covers the content, a topic not discussed in this work, but it also has the ability to over-emphasize or under-report certain events with little accountability.
\vspace{0.2cm}

This influence on coverage can be observed when interpreting the agglomeration dynamics highlighted by Fig.~\ref{fig:scatter_density}. Information sinks can be highlighted through the discovery of attraction poles in the embedding space.\\



Such sets of highly accretive sources, i.e. sources that draw other sources to align with their coverage, cluster neatly into large broadcast entities, some of which have been mentioned before. Fig.~\ref{fig:scatter_density}-(right) presents these groupings more exhaustively. The three large media groups chosen for analysis in Fig.~\ref{fig:flipbook} are present (\textit{Gray Television Inc.}, \textit{Sinclair Broadcasting Group} and \textit{GateHouse Media Inc.}), along with several other large players in the American media landscape.
\vspace{0.2cm}

Previous studies ~\cite{SemValk2000} and ~\cite{baldwin2010} have studied differences in terms of the types of content covered in television and newspapers outlets, finding that TV stations cover proportionally more ``global'' news than newspapers. Television is traditionally thought to be more impacted by media consolidation for this reason: content is costly to produce, hence it makes sense for large entities to share their footage at scale. In our model, this should intuitively lead to the flagging of television conglomerates as the strongest attractors, with high content similarities. However, we observe in Fig.~\ref{fig:scatter_density} that all mediums are represented and impacted by the convergence phenomenon. This could hint to the effect a convergence of mediums can have on the media landscape, with the efficacy co-ownership regulations being jeopardized by the all-encompassing nature of online content delivery.

\subsection{Interpretation of the temporal consistency}
\vspace{0.2cm}

None of this qualitative analysis would be possible without a temporal consistency constraint on the embedding space. Without such stability, the model could take advantage of an unnecessarily large number of degrees of freedom to align sources. In consequence, it would converge to very different solutions from one epoch to another. Due to the stochastic nature of the procedure, coverage changes would be rendered indistinguishable from optimization artifacts (see Fig.~\ref{tab:scores}). By penalizing sources that deviate from their previous positions, only significant coverage changes can force a source to migrate to a different region in space. In other words, in order to provoke a displacement, the channels' coverage should differ enough from the previous time step to outweigh the temporal constraint. If this condition is met, the source will converge towards a different neighborhood that better fits its coverage patterns, typically getting closer to similar channels.
\vspace{0.2cm}


This variability in time can be tuned through the regularization parameter $\lambda_{T}$, as detailed in Section~\ref{sec:model}, providing a way to highlight more global dynamics - in the case of strong regularization - or more individual variations - with weak regularization. We also observe that the constraint provides predictive gains. This can be explained by the accumulation of knowledge about sources over time. This last hypothesis is corroborated by the pattern observed in Fig.~\ref{fig:auc_time}, in which the accuracy reaches its maximum after the first few epochs before stabilizing until the end of the considered period. 
\vspace{0.2cm}


\section{Conclusion}
\label{sec:conclusion}
\vspace{0.2cm}

This work tackles the problem of dynamically modeling the filtering decisions of individual actors of the news ecosystem. Beyond the predictive capabilities of the approach, the knowledge gathered by the model is leveraged to characterize the evolution of the media landscape over time. Firstly, we report performance benefits of adding a temporal component to a coverage prediction model. In particular, we show that a dynamic embedding model is able to outperform existing approaches thanks to its ability to propagate knowledge obtained from former time slices to the current prediction step. Secondly, we demonstrate the application of this model as a framework in which to reason about the latent structure of media landscape, modeling the temporal evolution of news outlets' decision processes. Maintaining a consistent latent representation of sources' preferences enables powerful interpretation and visualization methods, highly effective in investigating shifts in the media ecosystem, at a large scale but also at the individual source level. 


We demonstrate the potential of the method on several channel acquisition campaigns. We show drastic post-acquisition content alignment in channels belonging to large, well-known broadcast conglomerates. This corroborates the hypothesis of deep consolidation of broadcast material inside news networks. Our work highlights the fragility of the external plurality assumption, which is supposed to guarantee a diversity of ownership and hence viewpoints. Finally, we automate this investigative process and explore several strategies to systemically identify abrupt variations in the news ecosystem, fingerprints of sharp changes in media programming.\\

\xhdr{Future work: } The main focus of this work was to provide a clear and insightful representation of the news landscape, and we foresee several directions to pursue this research effort. Firstly, the method is ripe for the addition of content-based refinements. Our approach focuses on the information's propagation patterns but it is still blind to the content itself and the way it is handled by news outlets. We could for example mention in this vein a semantic analysis of the covered articles. The fact that the model is content-agnostic can however be of great use for its application to other domains. Such methods could be applied to other dynamic systems in which information propagates and evolves, such as online social-networks, knowledge bases or citation networks. Lastly, as has been highlighted throughout this work, particularly in the context of the case-study, we believe that our work would strongly benefit from an interdisciplinary approach, providing tools to journalists, policy-makers or economists, whose expertise would also add great insight to our analysis.

\newpage
\bibliographystyle{ACM-Reference-Format}
\balance
\bibliography{refs}


\begin{thebibliography}{00}


\ifx \showCODEN    \undefined \def \showCODEN     #1{\unskip}     \fi
\ifx \showDOI      \undefined \def \showDOI       #1{#1}\fi
\ifx \showISBNx    \undefined \def \showISBNx     #1{\unskip}     \fi
\ifx \showISBNxiii \undefined \def \showISBNxiii  #1{\unskip}     \fi
\ifx \showISSN     \undefined \def \showISSN      #1{\unskip}     \fi
\ifx \showLCCN     \undefined \def \showLCCN      #1{\unskip}     \fi
\ifx \shownote     \undefined \def \shownote      #1{#1}          \fi
\ifx \showarticletitle \undefined \def \showarticletitle #1{#1}   \fi
\ifx \showURL      \undefined \def \showURL       {\relax}        \fi
\providecommand\bibfield[2]{#2}
\providecommand\bibinfo[2]{#2}
\providecommand\natexlab[1]{#1}
\providecommand\showeprint[2][]{arXiv:#2}

\bibitem[\protect\citeauthoryear{Acar, Dunlavy, and Kolda}{Acar
  et~al\mbox{.}}{2009}]%
        {acar2009link}
\bibfield{author}{\bibinfo{person}{Evrim Acar}, \bibinfo{person}{Daniel~M
  Dunlavy}, {and} \bibinfo{person}{Tamara~G Kolda}.}
  \bibinfo{year}{2009}\natexlab{}.
\newblock \showarticletitle{Link prediction on evolving data using matrix and
  tensor factorizations}. In \bibinfo{booktitle}{{\em Data Mining Workshops,
  2009. ICDMW'09. IEEE International Conference on}}. IEEE,
  \bibinfo{pages}{262--269}.
\newblock


\bibitem[\protect\citeauthoryear{An and Kwak}{An and Kwak}{2016}]%
        {an2016two}
\bibfield{author}{\bibinfo{person}{Jisun An} {and} \bibinfo{person}{Haewon
  Kwak}.} \bibinfo{year}{2016}\natexlab{}.
\newblock \showarticletitle{Two Tales of the World: Comparison of Widely Used
  World News Datasets GDELT and EventRegistry}. In \bibinfo{booktitle}{{\em
  Proceedings of the 10th International AAAI Conference on Web and Social Media
  (ICWSM16)}}. \bibinfo{pages}{619--622}.
\newblock


\bibitem[\protect\citeauthoryear{Anderson and Coate}{Anderson and
  Coate}{2005}]%
        {10.2307/3700696}
\bibfield{author}{\bibinfo{person}{Simon~P. Anderson} {and}
  \bibinfo{person}{Stephen Coate}.} \bibinfo{year}{2005}\natexlab{}.
\newblock \showarticletitle{Market Provision of Broadcasting: A Welfare
  Analysis}.
\newblock \bibinfo{journal}{{\em The Review of Economic Studies\/}}
  \bibinfo{volume}{72}, \bibinfo{number}{4} (\bibinfo{year}{2005}),
  \bibinfo{pages}{947--972}.
\newblock
\showISSN{00346527, 1467937X}
\showURL{%
\url{http://www.jstor.org/stable/3700696}}


\bibitem[\protect\citeauthoryear{Baldwin}{Baldwin}{2010}]%
        {baldwin2010}
\bibfield{author}{\bibinfo{person}{Bergan D. Fico F. Lacy S. Wildman~S.S.
  Baldwin, T.}} \bibinfo{year}{2010}\natexlab{}.
\newblock \showarticletitle{News Media Coverage of City Governments in 2009}.
\newblock  (\bibinfo{year}{2010}).
\newblock


\bibitem[\protect\citeauthoryear{Bourgeois, Rappaz, and Aberer}{Bourgeois
  et~al\mbox{.}}{2018}]%
        {bourgeois2018selection}
\bibfield{author}{\bibinfo{person}{Dylan Bourgeois},
  \bibinfo{person}{J{\'e}r{\'e}mie Rappaz}, {and} \bibinfo{person}{Karl
  Aberer}.} \bibinfo{year}{2018}\natexlab{}.
\newblock \showarticletitle{Selection Bias in News Coverage: Learning it,
  Fighting it}. In \bibinfo{booktitle}{{\em The International World Wide Web
  Conference 2018}}.
\newblock


\bibitem[\protect\citeauthoryear{DellaVigna, Kaplan, Krueger, Manacorda,
  Moretti, Persson, Popkin, Puglisi, Rabin, Shapiro, Simonsohn, Stoker,
  Stromberg, Deryugina, Deza, Fox, Galicia, Ho, Khanchanawong, Kim, Kohan,
  Kumar, Leung, Li, Lim, Mai, Parekh, Radakrishnan, Relan, Acland, Bhargava,
  Ebenstein, and Pope}{DellaVigna et~al\mbox{.}}{2005}]%
        {DellaVigna2005TheFN}
\bibfield{author}{\bibinfo{person}{Stefano DellaVigna}, \bibinfo{person}{Ethan
  Kaplan}, \bibinfo{person}{Alan B~. Krueger}, \bibinfo{person}{Marco
  Manacorda}, \bibinfo{person}{Enrico Moretti}, \bibinfo{person}{Torsten
  Persson}, \bibinfo{person}{Sam Popkin}, \bibinfo{person}{Riccardo Puglisi},
  \bibinfo{person}{Matthew Rabin}, \bibinfo{person}{Jesse~M. Shapiro},
  \bibinfo{person}{Uri Simonsohn}, \bibinfo{person}{Laura Stoker},
  \bibinfo{person}{David Stromberg}, \bibinfo{person}{Tatyana Deryugina},
  \bibinfo{person}{Monica Deza}, \bibinfo{person}{Dylan Fox},
  \bibinfo{person}{Melissa Galicia}, \bibinfo{person}{Calvin Wai-Loon Ho},
  \bibinfo{person}{Sudhamas Khanchanawong}, \bibinfo{person}{Richard~M. Kim},
  \bibinfo{person}{Martin Kohan}, \bibinfo{person}{Vipul~Surender Kumar},
  \bibinfo{person}{Jonathan~J. Leung}, \bibinfo{person}{Clarice Li},
  \bibinfo{person}{Tze~Yang Lim}, \bibinfo{person}{Ming Mai},
  \bibinfo{person}{Sameer Parekh}, \bibinfo{person}{Sharmini Radakrishnan},
  \bibinfo{person}{Rohan Relan}, \bibinfo{person}{Dan Acland},
  \bibinfo{person}{Saurabh Bhargava}, \bibinfo{person}{Avi Ebenstein}, {and}
  \bibinfo{person}{Devin~G. Pope}.} \bibinfo{year}{2005}\natexlab{}.
\newblock \showarticletitle{The Fox News Effect: Media Bias and Voting}.
\newblock


\bibitem[\protect\citeauthoryear{Djankov, McLiesh, Nenova, and
  Shleifer}{Djankov et~al\mbox{.}}{2003}]%
        {Djankov03}
\bibfield{author}{\bibinfo{person}{Simeon Djankov}, \bibinfo{person}{Caralee
  McLiesh}, \bibinfo{person}{Tatiana Nenova}, {and} \bibinfo{person}{Andrei
  Shleifer}.} \bibinfo{year}{2003}\natexlab{}.
\newblock \showarticletitle{Who Owns the Media?}
\newblock \bibinfo{journal}{{\em Journal of Law and Economics\/}}
  \bibinfo{volume}{46}, \bibinfo{number}{2} (\bibinfo{year}{2003}),
  \bibinfo{pages}{341--381}.
\newblock


\bibitem[\protect\citeauthoryear{Doyle}{Doyle}{2002}]%
        {mediaconcentration}
\bibfield{author}{\bibinfo{person}{Gillian Doyle}.}
  \bibinfo{year}{2002}\natexlab{}.
\newblock \bibinfo{booktitle}{{\em Media Ownership: The Economics and Politics
  of Convergence and Concentration in the UK and European Media}}.
\newblock
\showDOI{%
\url{https://doi.org/10.4135/9781446219942}}


\bibitem[\protect\citeauthoryear{Dunlavy, Kolda, and Acar}{Dunlavy
  et~al\mbox{.}}{2011}]%
        {dunlavy2011temporal}
\bibfield{author}{\bibinfo{person}{Daniel~M Dunlavy}, \bibinfo{person}{Tamara~G
  Kolda}, {and} \bibinfo{person}{Evrim Acar}.} \bibinfo{year}{2011}\natexlab{}.
\newblock \showarticletitle{Temporal link prediction using matrix and tensor
  factorizations}.
\newblock \bibinfo{journal}{{\em ACM Transactions on Knowledge Discovery from
  Data (TKDD)\/}} \bibinfo{volume}{5}, \bibinfo{number}{2}
  (\bibinfo{year}{2011}), \bibinfo{pages}{10}.
\newblock


\bibitem[\protect\citeauthoryear{Gentzkow and Shapiro}{Gentzkow and
  Shapiro}{2010}]%
        {RePEc:ecm:emetrp:v:78:y:2010:i:1:p:35-71}
\bibfield{author}{\bibinfo{person}{Matthew Gentzkow} {and}
  \bibinfo{person}{Jesse Shapiro}.} \bibinfo{year}{2010}\natexlab{}.
\newblock \showarticletitle{What Drives Media Slant? Evidence From U.S. Daily
  Newspapers}.
\newblock \bibinfo{journal}{{\em Econometrica\/}} \bibinfo{volume}{78},
  \bibinfo{number}{1} (\bibinfo{year}{2010}), \bibinfo{pages}{35--71}.
\newblock
\showURL{%
\url{https://EconPapers.repec.org/RePEc:ecm:emetrp:v:78:y:2010:i:1:p:35-71}}


\bibitem[\protect\citeauthoryear{Gentzkow, Shapiro, and Sinkinson}{Gentzkow
  et~al\mbox{.}}{2009}]%
        {NBERw15544}
\bibfield{author}{\bibinfo{person}{Matthew Gentzkow}, \bibinfo{person}{Jesse~M
  Shapiro}, {and} \bibinfo{person}{Michael Sinkinson}.}
  \bibinfo{year}{2009}\natexlab{}.
\newblock \bibinfo{booktitle}{{\em The Effect of Newspaper Entry and Exit on
  Electoral Politics}}.
\newblock \bibinfo{type}{Working Paper} 15544. \bibinfo{institution}{National
  Bureau of Economic Research}.
\newblock
\showDOI{%
\url{https://doi.org/10.3386/w15544}}


\bibitem[\protect\citeauthoryear{George and Waldfogel}{George and
  Waldfogel}{2008}]%
        {georgewald08}
\bibfield{author}{\bibinfo{person}{L.M. George} {and} \bibinfo{person}{Joel
  Waldfogel}.} \bibinfo{year}{2008}\natexlab{}.
\newblock \showarticletitle{National Media and Local Political Participation:
  The Case of the New York Times}.
\newblock  (\bibinfo{date}{01} \bibinfo{year}{2008}), \bibinfo{pages}{33--48}.
\newblock


\bibitem[\protect\citeauthoryear{Gerner, Abu-Jabr, Schrodt, and Ömür
  Yilmaz}{Gerner et~al\mbox{.}}{2002}]%
        {Gerner02conflictand}
\bibfield{author}{\bibinfo{person}{Deborah~J. Gerner}, \bibinfo{person}{Rajaa
  Abu-Jabr}, \bibinfo{person}{Philip~A. Schrodt}, {and} \bibinfo{person}{Ömür
  Yilmaz}.} \bibinfo{year}{2002}\natexlab{}.
\newblock \showarticletitle{Conflict and Mediation Event Observations (CAMEO):
  A New Event Data Framework for the Analysis of Foreign Policy Interactions}.
  In \bibinfo{booktitle}{{\em of Foreign Policy Interactions. Paper presented
  at the International Studies Association}}.
\newblock


\bibitem[\protect\citeauthoryear{Gleditsch, Metternich, and Ruggeri}{Gleditsch
  et~al\mbox{.}}{2014}]%
        {doi:10.1177/0022343313496803}
\bibfield{author}{\bibinfo{person}{Kristian~Skrede Gleditsch},
  \bibinfo{person}{Nils~W Metternich}, {and} \bibinfo{person}{Andrea Ruggeri}.}
  \bibinfo{year}{2014}\natexlab{}.
\newblock \showarticletitle{Data and progress in peace and conflict research}.
\newblock \bibinfo{journal}{{\em Journal of Peace Research\/}}
  \bibinfo{volume}{51}, \bibinfo{number}{2} (\bibinfo{year}{2014}),
  \bibinfo{pages}{301--314}.
\newblock
\showDOI{%
\url{https://doi.org/10.1177/0022343313496803}}
\showeprint{https://doi.org/10.1177/0022343313496803}


\bibitem[\protect\citeauthoryear{Goldstein}{Goldstein}{1992}]%
        {10.2307/174480}
\bibfield{author}{\bibinfo{person}{Joshua~S. Goldstein}.}
  \bibinfo{year}{1992}\natexlab{}.
\newblock \showarticletitle{A Conflict-Cooperation Scale for WEIS Events Data}.
\newblock \bibinfo{journal}{{\em The Journal of Conflict Resolution\/}}
  \bibinfo{volume}{36}, \bibinfo{number}{2} (\bibinfo{year}{1992}),
  \bibinfo{pages}{369--385}.
\newblock
\showISSN{00220027, 15528766}
\showURL{%
\url{http://www.jstor.org/stable/174480}}


\bibitem[\protect\citeauthoryear{Groseclose and Milyo}{Groseclose and
  Milyo}{2005}]%
        {RePEc:oup:qjecon:v:120:y:2005:i:4:p:1191-1237.}
\bibfield{author}{\bibinfo{person}{Tim Groseclose} {and}
  \bibinfo{person}{Jeffrey Milyo}.} \bibinfo{year}{2005}\natexlab{}.
\newblock \showarticletitle{A Measure of Media Bias}.
\newblock \bibinfo{journal}{{\em The Quarterly Journal of Economics\/}}
  \bibinfo{volume}{120}, \bibinfo{number}{4} (\bibinfo{year}{2005}),
  \bibinfo{pages}{1191--1237}.
\newblock
\showURL{%
\url{https://EconPapers.repec.org/RePEc:oup:qjecon:v:120:y:2005:i:4:p:1191-1237.}}


\bibitem[\protect\citeauthoryear{He, Fang, Wang, and McAuley}{He
  et~al\mbox{.}}{2016}]%
        {he2016vista}
\bibfield{author}{\bibinfo{person}{Ruining He}, \bibinfo{person}{Chen Fang},
  \bibinfo{person}{Zhaowen Wang}, {and} \bibinfo{person}{Julian McAuley}.}
  \bibinfo{year}{2016}\natexlab{}.
\newblock \showarticletitle{Vista: a visually, socially, and temporally-aware
  model for artistic recommendation}. In \bibinfo{booktitle}{{\em Proceedings
  of the 10th ACM Conference on Recommender Systems}}. ACM,
  \bibinfo{pages}{309--316}.
\newblock


\bibitem[\protect\citeauthoryear{He and McAuley}{He and McAuley}{2016}]%
        {he2016ups}
\bibfield{author}{\bibinfo{person}{Ruining He} {and} \bibinfo{person}{Julian
  McAuley}.} \bibinfo{year}{2016}\natexlab{}.
\newblock \showarticletitle{Ups and downs: Modeling the visual evolution of
  fashion trends with one-class collaborative filtering}. In
  \bibinfo{booktitle}{{\em proceedings of the 25th international conference on
  world wide web}}. International World Wide Web Conferences Steering
  Committee, \bibinfo{pages}{507--517}.
\newblock


\bibitem[\protect\citeauthoryear{Hu, Koren, and Volinsky}{Hu
  et~al\mbox{.}}{2008}]%
        {Hu2008CollaborativeFF}
\bibfield{author}{\bibinfo{person}{Yifan Hu}, \bibinfo{person}{Yehuda Koren},
  {and} \bibinfo{person}{Chris Volinsky}.} \bibinfo{year}{2008}\natexlab{}.
\newblock \showarticletitle{Collaborative Filtering for Implicit Feedback
  Datasets}.
\newblock \bibinfo{journal}{{\em 2008 Eighth IEEE International Conference on
  Data Mining\/}} (\bibinfo{year}{2008}), \bibinfo{pages}{263--272}.
\newblock


\bibitem[\protect\citeauthoryear{Jenkins}{Jenkins}{2004}]%
        {jenkins2004cultural}
\bibfield{author}{\bibinfo{person}{Henry Jenkins}.}
  \bibinfo{year}{2004}\natexlab{}.
\newblock \showarticletitle{The cultural logic of media convergence}.
\newblock \bibinfo{journal}{{\em International journal of cultural studies\/}}
  \bibinfo{volume}{7}, \bibinfo{number}{1} (\bibinfo{year}{2004}),
  \bibinfo{pages}{33--43}.
\newblock


\bibitem[\protect\citeauthoryear{Keertipati, Savarimuthu, Purvis, and
  Purvis}{Keertipati et~al\mbox{.}}{2014}]%
        {Keertipati2014MultilevelAO}
\bibfield{author}{\bibinfo{person}{Swetha Keertipati}, \bibinfo{person}{Bastin
  Tony~Roy Savarimuthu}, \bibinfo{person}{Maryam Purvis}, {and}
  \bibinfo{person}{Martin~K. Purvis}.} \bibinfo{year}{2014}\natexlab{}.
\newblock \showarticletitle{Multi-level Analysis of Peace and Conflict Data in
  GDELT}. In \bibinfo{booktitle}{{\em MLSDA@PRICAI}}.
\newblock


\bibitem[\protect\citeauthoryear{Koren}{Koren}{2009}]%
        {koren2009collaborative}
\bibfield{author}{\bibinfo{person}{Yehuda Koren}.}
  \bibinfo{year}{2009}\natexlab{}.
\newblock \showarticletitle{Collaborative filtering with temporal dynamics}. In
  \bibinfo{booktitle}{{\em Proceedings of the 15th ACM SIGKDD international
  conference on Knowledge discovery and data mining}}. ACM,
  \bibinfo{pages}{447--456}.
\newblock


\bibitem[\protect\citeauthoryear{Kwak and An}{Kwak and An}{2014}]%
        {kwak2014first}
\bibfield{author}{\bibinfo{person}{Haewoon Kwak} {and} \bibinfo{person}{Jisun
  An}.} \bibinfo{year}{2014}\natexlab{}.
\newblock \showarticletitle{A first look at global news coverage of disasters
  by using the gdelt dataset}. In \bibinfo{booktitle}{{\em International
  Conference on Social Informatics}}. Springer, \bibinfo{pages}{300--308}.
\newblock


\bibitem[\protect\citeauthoryear{Lahoti, Garimella, and Gionis}{Lahoti
  et~al\mbox{.}}{2018}]%
        {lahoti2018joint}
\bibfield{author}{\bibinfo{person}{Preethi Lahoti}, \bibinfo{person}{Kiran
  Garimella}, {and} \bibinfo{person}{Aristides Gionis}.}
  \bibinfo{year}{2018}\natexlab{}.
\newblock \showarticletitle{Joint non-negative matrix factorization for
  learning ideological leaning on Twitter}. In \bibinfo{booktitle}{{\em
  Proceedings of the Eleventh ACM International Conference on Web Search and
  Data Mining}}. ACM, \bibinfo{pages}{351--359}.
\newblock


\bibitem[\protect\citeauthoryear{Laurance}{Laurance}{1990}]%
        {Laurance1990}
\bibfield{author}{\bibinfo{person}{Edward~J. Laurance}.}
  \bibinfo{year}{1990}\natexlab{}.
\newblock \showarticletitle{Events data and policy analysis:}.
\newblock \bibinfo{journal}{{\em Policy Sciences\/}} \bibinfo{volume}{23},
  \bibinfo{number}{2} (\bibinfo{date}{01 May} \bibinfo{year}{1990}),
  \bibinfo{pages}{111--132}.
\newblock
\showISSN{1573-0891}
\showDOI{%
\url{https://doi.org/10.1007/BF00175597}}


\bibitem[\protect\citeauthoryear{Leetaru and Schrodt}{Leetaru and
  Schrodt}{2013}]%
        {Leetaru13gdelt:global}
\bibfield{author}{\bibinfo{person}{Kalev Leetaru} {and}
  \bibinfo{person}{Philip~A. Schrodt}.} \bibinfo{year}{2013}\natexlab{}.
\newblock \showarticletitle{GDELT: Global data on events, location, and tone}.
\newblock \bibinfo{journal}{{\em ISA Annual Convention\/}}
  (\bibinfo{year}{2013}).
\newblock


\bibitem[\protect\citeauthoryear{Mondak}{Mondak}{1995}]%
        {mondak95}
\bibfield{author}{\bibinfo{person}{{Jeffery J.} Mondak}.}
  \bibinfo{year}{1995}\natexlab{}.
\newblock \showarticletitle{Media exposure and political discussion in u.s.
  elections}.
\newblock \bibinfo{journal}{{\em Journal of Politics\/}} \bibinfo{volume}{57},
  \bibinfo{number}{1} (\bibinfo{date}{1 5} \bibinfo{year}{1995}),
  \bibinfo{pages}{62--85}.
\newblock
\showISSN{0022-3816}
\showDOI{%
\url{https://doi.org/10.2307/2960271}}


\bibitem[\protect\citeauthoryear{Oberholzer-Gee and Waldfogel}{Oberholzer-Gee
  and Waldfogel}{2006}]%
        {NBERw12317}
\bibfield{author}{\bibinfo{person}{Felix Oberholzer-Gee} {and}
  \bibinfo{person}{Joel Waldfogel}.} \bibinfo{year}{2006}\natexlab{}.
\newblock \bibinfo{booktitle}{{\em Media Markets and Localism: Does Local News
  en Español Boost Hispanic Voter Turnout?}}
\newblock \bibinfo{type}{Working Paper} 12317. \bibinfo{institution}{National
  Bureau of Economic Research}.
\newblock
\showDOI{%
\url{https://doi.org/10.3386/w12317}}


\bibitem[\protect\citeauthoryear{Olteanu, Castillo, Diakopoulos, and
  Aberer}{Olteanu et~al\mbox{.}}{2015}]%
        {OlteanuCDA15}
\bibfield{author}{\bibinfo{person}{Alexandra Olteanu}, \bibinfo{person}{Carlos
  Castillo}, \bibinfo{person}{Nicholas Diakopoulos}, {and}
  \bibinfo{person}{Karl Aberer}.} \bibinfo{year}{2015}\natexlab{}.
\newblock \showarticletitle{Comparing Events Coverage in Online News and Social
  Media: The Case of Climate Change}. In \bibinfo{booktitle}{{\em {ICWSM}}}.
  \bibinfo{publisher}{{AAAI} Press}, \bibinfo{pages}{288--297}.
\newblock


\bibitem[\protect\citeauthoryear{Pan, Zhou, Cao, Liu, Lukose, Scholz, and
  Yang}{Pan et~al\mbox{.}}{2008}]%
        {pan2008one}
\bibfield{author}{\bibinfo{person}{Rong Pan}, \bibinfo{person}{Yunhong Zhou},
  \bibinfo{person}{Bin Cao}, \bibinfo{person}{Nathan~N Liu},
  \bibinfo{person}{Rajan Lukose}, \bibinfo{person}{Martin Scholz}, {and}
  \bibinfo{person}{Qiang Yang}.} \bibinfo{year}{2008}\natexlab{}.
\newblock \showarticletitle{One-class collaborative filtering}. In
  \bibinfo{booktitle}{{\em Data Mining, 2008. ICDM'08. Eighth IEEE
  International Conference on}}. IEEE, \bibinfo{pages}{502--511}.
\newblock


\bibitem[\protect\citeauthoryear{Potter and Matsa}{Potter and Matsa}{2014}]%
        {pew2013}
\bibfield{author}{\bibinfo{person}{Deborah Potter} {and}
  \bibinfo{person}{Katerina~Eva Matsa}.} \bibinfo{year}{2014}\natexlab{}.
\newblock \bibinfo{booktitle}{{\em State of the News Media 2014: A Boom in
  Acquisitions and Content Sharing Shapes Local TV News in 2013}}.
\newblock \bibinfo{type}{{T}echnical {R}eport}. \bibinfo{institution}{Pew
  Research Center}.
\newblock


\bibitem[\protect\citeauthoryear{Pritchard}{Pritchard}{2002}]%
        {Pritchard}
\bibfield{author}{\bibinfo{person}{David Pritchard}.}
  \bibinfo{year}{2002}\natexlab{}.
\newblock \showarticletitle{Viewpoint Diversity in Cross-Owned Newspapers and
  Television Stations: A Study of News Coverage of the 2000 Presidential
  Campaign}.
\newblock  (\bibinfo{date}{September} \bibinfo{year}{2002}).
\newblock
\showURL{%
\url{https://docs.fcc.gov/public/attachments/DOC-226838A7.pdf}}


\bibitem[\protect\citeauthoryear{Qiao, Li, Deng, Ding, and Wang}{Qiao
  et~al\mbox{.}}{2015}]%
        {Qiao:2015:GMD:2859846.2860085}
\bibfield{author}{\bibinfo{person}{Fengcai Qiao}, \bibinfo{person}{Pei Li},
  \bibinfo{person}{Jingsheng Deng}, \bibinfo{person}{Zhaoyun Ding}, {and}
  \bibinfo{person}{Hui Wang}.} \bibinfo{year}{2015}\natexlab{}.
\newblock \showarticletitle{Graph-Based Method for Detecting Occupy Protest
  Events Using GDELT Dataset}. In \bibinfo{booktitle}{{\em Proceedings of the
  2015 International Conference on Cyber-Enabled Distributed Computing and
  Knowledge Discovery}} {\em (\bibinfo{series}{CYBERC '15})}.
  \bibinfo{publisher}{IEEE Computer Society}, \bibinfo{address}{Washington, DC,
  USA}, \bibinfo{pages}{164--168}.
\newblock
\showISBNx{978-1-4673-9200-6}
\showDOI{%
\url{https://doi.org/10.1109/CyberC.2015.77}}


\bibitem[\protect\citeauthoryear{Rendle and Freudenthaler}{Rendle and
  Freudenthaler}{2014}]%
        {rendle2014improving}
\bibfield{author}{\bibinfo{person}{Steffen Rendle} {and}
  \bibinfo{person}{Christoph Freudenthaler}.} \bibinfo{year}{2014}\natexlab{}.
\newblock \showarticletitle{Improving pairwise learning for item recommendation
  from implicit feedback}. In \bibinfo{booktitle}{{\em Proceedings of the 7th
  ACM international conference on Web search and data mining}}. ACM,
  \bibinfo{pages}{273--282}.
\newblock


\bibitem[\protect\citeauthoryear{Rendle, Freudenthaler, Gantner, and
  Schmidt-Thieme}{Rendle et~al\mbox{.}}{2009}]%
        {Rendle2009BPRBP}
\bibfield{author}{\bibinfo{person}{Steffen Rendle}, \bibinfo{person}{Christoph
  Freudenthaler}, \bibinfo{person}{Zeno Gantner}, {and} \bibinfo{person}{Lars
  Schmidt-Thieme}.} \bibinfo{year}{2009}\natexlab{}.
\newblock \showarticletitle{BPR: Bayesian Personalized Ranking from Implicit
  Feedback}. In \bibinfo{booktitle}{{\em UAI}}.
\newblock


\bibitem[\protect\citeauthoryear{Rudolph and Blei}{Rudolph and Blei}{2017}]%
        {rudolph2017dynamic}
\bibfield{author}{\bibinfo{person}{Maja Rudolph} {and} \bibinfo{person}{David
  Blei}.} \bibinfo{year}{2017}\natexlab{}.
\newblock \showarticletitle{Dynamic Bernoulli embeddings for language
  evolution}.
\newblock \bibinfo{journal}{{\em arXiv preprint arXiv:1703.08052\/}}
  (\bibinfo{year}{2017}).
\newblock


\bibitem[\protect\citeauthoryear{Rudolph, Ruiz, Mandt, and Blei}{Rudolph
  et~al\mbox{.}}{2016}]%
        {rudolph2016exponential}
\bibfield{author}{\bibinfo{person}{Maja Rudolph}, \bibinfo{person}{Francisco
  Ruiz}, \bibinfo{person}{Stephan Mandt}, {and} \bibinfo{person}{David Blei}.}
  \bibinfo{year}{2016}\natexlab{}.
\newblock \showarticletitle{Exponential family embeddings}. In
  \bibinfo{booktitle}{{\em Advances in Neural Information Processing Systems}}.
  \bibinfo{pages}{478--486}.
\newblock


\bibitem[\protect\citeauthoryear{Saez-Trumper, Castillo, and
  Lalmas}{Saez-Trumper et~al\mbox{.}}{2013}]%
        {saez2013social}
\bibfield{author}{\bibinfo{person}{Diego Saez-Trumper}, \bibinfo{person}{Carlos
  Castillo}, {and} \bibinfo{person}{Mounia Lalmas}.}
  \bibinfo{year}{2013}\natexlab{}.
\newblock \showarticletitle{Social media news communities: gatekeeping,
  coverage, and statement bias}. In \bibinfo{booktitle}{{\em Proceedings of the
  22nd ACM international conference on Conference on information \& knowledge
  management}}. ACM, \bibinfo{pages}{1679--1684}.
\newblock


\bibitem[\protect\citeauthoryear{S{\'a}ez-Trumper, Castillo, and
  Lalmas}{S{\'a}ez-Trumper et~al\mbox{.}}{2013}]%
        {SezTrumper2013SocialMN}
\bibfield{author}{\bibinfo{person}{Diego S{\'a}ez-Trumper},
  \bibinfo{person}{Carlos Castillo}, {and} \bibinfo{person}{Mounia Lalmas}.}
  \bibinfo{year}{2013}\natexlab{}.
\newblock \showarticletitle{Social media news communities: gatekeeping,
  coverage, and statement bias}. In \bibinfo{booktitle}{{\em CIKM}}.
\newblock


\bibitem[\protect\citeauthoryear{Schrodt, Davis, and Weddle}{Schrodt
  et~al\mbox{.}}{1994}]%
        {schrodt1994political}
\bibfield{author}{\bibinfo{person}{P.~A. Schrodt}, \bibinfo{person}{S.~G.
  Davis}, {and} \bibinfo{person}{J.~L. Weddle}.}
  \bibinfo{year}{1994}\natexlab{}.
\newblock \showarticletitle{{Political Science: KEDS--A Program for the Machine
  Coding of Event Data}}.
\newblock \bibinfo{journal}{{\em Social Science Computer Review\/}}
  \bibinfo{volume}{12}, \bibinfo{number}{4} (\bibinfo{year}{1994}),
  \bibinfo{pages}{561}.
\newblock


\bibitem[\protect\citeauthoryear{Semetko and Valkenburg}{Semetko and
  Valkenburg}{2000}]%
        {SemValk2000}
\bibfield{author}{\bibinfo{person}{Holli Semetko} {and} \bibinfo{person}{Patti
  Valkenburg}.} \bibinfo{year}{2000}\natexlab{}.
\newblock \showarticletitle{Framing European Politics: A Content Analysis of
  Press and Television News}.
\newblock   \bibinfo{volume}{50} (\bibinfo{date}{06} \bibinfo{year}{2000}),
  \bibinfo{pages}{93 -- 109}.
\newblock


\bibitem[\protect\citeauthoryear{Silverman}{Silverman}{1986}]%
        {Silverman86}
\bibfield{author}{\bibinfo{person}{B.~W. Silverman}.}
  \bibinfo{year}{1986}\natexlab{}.
\newblock \bibinfo{booktitle}{{\em Density Estimation for Statistics and Data
  Analysis}}.
\newblock \bibinfo{publisher}{Chapman \& Hall}, \bibinfo{address}{London}.
\newblock


\bibitem[\protect\citeauthoryear{Steiner}{Steiner}{1952}]%
        {RePEc:oup:qjecon:v:66:y:1952:i:2:p:194-223.}
\bibfield{author}{\bibinfo{person}{Peter~O. Steiner}.}
  \bibinfo{year}{1952}\natexlab{}.
\newblock \showarticletitle{Program Patterns and Preferences, and the
  Workability of Competition in Radio Broadcasting}.
\newblock \bibinfo{journal}{{\em The Quarterly Journal of Economics\/}}
  \bibinfo{volume}{66}, \bibinfo{number}{2} (\bibinfo{year}{1952}),
  \bibinfo{pages}{194--223}.
\newblock
\showURL{%
\url{https://EconPapers.repec.org/RePEc:oup:qjecon:v:66:y:1952:i:2:p:194-223.}}


\bibitem[\protect\citeauthoryear{van~der Maaten and Hinton}{van~der Maaten and
  Hinton}{2008}]%
        {tsne}
\bibfield{author}{\bibinfo{person}{L.J.P van~der Maaten} {and}
  \bibinfo{person}{G.E. Hinton}.} \bibinfo{year}{Nov 2008}\natexlab{}.
\newblock \showarticletitle{Visualizing High-Dimensional Data Using t-SNE}.
\newblock \bibinfo{journal}{{\em Journal of Machine Learning Research\/}}
  \bibinfo{volume}{9: 2579–2605} (\bibinfo{year}{Nov 2008}).
\newblock


\bibitem[\protect\citeauthoryear{Vizcarrondo}{Vizcarrondo}{2013}]%
        {vizcarrondo2013measuring}
\bibfield{author}{\bibinfo{person}{Tom Vizcarrondo}.}
  \bibinfo{year}{2013}\natexlab{}.
\newblock \showarticletitle{Measuring concentration of media ownership:
  1976--2009}.
\newblock \bibinfo{journal}{{\em International Journal on Media Management\/}}
  \bibinfo{volume}{15}, \bibinfo{number}{3} (\bibinfo{year}{2013}),
  \bibinfo{pages}{177--195}.
\newblock


\bibitem[\protect\citeauthoryear{Xiong, Chen, Huang, Schneider, and
  Carbonell}{Xiong et~al\mbox{.}}{2010}]%
        {xiong2010temporal}
\bibfield{author}{\bibinfo{person}{Liang Xiong}, \bibinfo{person}{Xi Chen},
  \bibinfo{person}{Tzu-Kuo Huang}, \bibinfo{person}{Jeff Schneider}, {and}
  \bibinfo{person}{Jaime~G Carbonell}.} \bibinfo{year}{2010}\natexlab{}.
\newblock \showarticletitle{Temporal collaborative filtering with bayesian
  probabilistic tensor factorization}. In \bibinfo{booktitle}{{\em Proceedings
  of the 2010 SIAM International Conference on Data Mining}}. SIAM,
  \bibinfo{pages}{211--222}.
\newblock


\bibitem[\protect\citeauthoryear{Yonamine}{Yonamine}{2013}]%
        {yonamine2013nuanced}
\bibfield{author}{\bibinfo{person}{James~Edward Yonamine}.}
  \bibinfo{year}{2013}\natexlab{}.
\newblock \showarticletitle{A nuanced study of political conflict using the
  global datasets of events location and tone (GDELT) dataset}.
\newblock  (\bibinfo{year}{2013}).
\newblock


\bibitem[\protect\citeauthoryear{Yu, Aggarwal, and Wang}{Yu
  et~al\mbox{.}}{2017}]%
        {yu2017temporally}
\bibfield{author}{\bibinfo{person}{Wenchao Yu}, \bibinfo{person}{Charu~C
  Aggarwal}, {and} \bibinfo{person}{Wei Wang}.}
  \bibinfo{year}{2017}\natexlab{}.
\newblock \showarticletitle{Temporally factorized network modeling for
  evolutionary network analysis}. In \bibinfo{booktitle}{{\em Proceedings of
  the Tenth ACM International Conference on Web Search and Data Mining}}. ACM,
  \bibinfo{pages}{455--464}.
\newblock


\end{thebibliography}

\end{document}